\documentstyle[psfig,twocolumn,aps]{revtex}
\begin{document}
%\eqnobysec   %JPB
%\jl{2}    %JPB
%\twocolumn[
\draft    %JPB

\title{Atom loss and the formation of a molecular Bose-Einstein
condensate by Feshbach resonance}

\author{V. A. Yurovsky, A. Ben-Reuven,}

\address{School of Chemistry, Tel Aviv University, 69978
Tel Aviv, Israel}

\author{P. S. Julienne and C. J. Williams}

\address{Atomic Physics Division, Stop 8423, National
Institute of Standards and Technology, Gaithersburg, MD 20889}

\date{\today}
\maketitle    %JPB
%\widetext

\begin{abstract}  In experiments conducted recently at MIT
on Na Bose-Einstein condensates [S. Inouye {\it et al}, Nature
{\bf 392}, 151 (1998);  J. Stenger {\it et al}, Phys. Rev. Lett.
{\bf 82}, 2422 (1999)], large loss rates were observed when a
time-varying magnetic field was used to tune a molecular
Feshbach resonance state near the state of a pair  of atoms in
the condensate. A collisional deactivation mechanism affecting a
temporarily formed molecular condensate [see V. A. Yurovsky, A.
Ben-Reuven, P. S. Julienne and C. J. Williams, Phys. Rev. A {\bf
60}, R765 (1999)], studied here in more detail, accounts for the
results of the slow-sweep experiments. A best fit to the MIT
data yields a rate coefficient for deactivating atom-molecule
collisions of $1.6\times 10^{-10}$  cm$^{3}/$s. In the case of
the fast sweep experiment, a study is carried out of the
combined effect of two competing mechanisms, the three-atom
(atom-molecule) or four-atom (molecule-molecule) collisional
deactivation vs. a process of two-atom trap-state excitation by
curve crossing [F. H. Mies, P. S. Julienne, and E. Tiesinga,
Phys. Rev. A {\bf 61}, 022721 (2000)]. It is shown that both
mechanisms contribute to the loss comparably and nonadditively.

\end{abstract}
%]
\narrowtext

\section{Introduction} \label{Introd}

Most properties of a Bose-Einstein condensate (BEC) are
determined by interatomic interactions (see Refs.\
\cite{PW98,DGPS99}). These interactions are responsible for the
characteristic nonlinear term in the equations of motion of the
condensate, whose magnitude depends on the collisional elastic
scattering length. Recent experiments \cite{IASMSK98,SIAMSK99} on
optically-trapped BECs drew attention to the effects of Feshbach
resonances on the properties of the condensate, as scattering lengths
are strongly modified by the presence of a resonance. More
particularly, these experiments sought to modify these effects by
application of a magnetic field (see Refs.\ \cite{Verhaar,TTHK99}).
One of the rather astonishing results observed was a dramatic loss of
condensate population as the magnetic field was varied so that the
ensuing Zeeman shift made the system pass through a resonance, or
approach it closely.

A Feshbach resonance may exist when the energy of a pair of
atoms in the condensate is close to that of a metastable molecular
state Na$_{2}\left( m\right) $. Then the scattering length varies
 strongly as a
function of the energy mismatch between the two states. This mismatch
can be controlled by applying a varying magnetic field. The energies
of the two states can be brought closer to each other, as the two
states have different Zeeman shifts. The MIT experiment
\cite{IASMSK98,SIAMSK99} strived to study the effect of the Zeeman
shift by applying a time-varying magnetic field in two distinct
procedures: (a) a fast sweep through the resonance, using a fast ramp
speed of the magnetic field, and (b) a slow sweep, using much lower
speeds, in which the ramp is stopped short of crossing the resonance.
The latter procedure is then repeated by using different values of
the stopping value of the magnetic field, and is carried out on both
sides of the resonance. Both types of experiment resulted in a large
condensate population loss.

 This work is devoted to the study of possible mechanisms
leading to this loss. Preliminary results were presented in Refs.\
\cite{YBJW99a,YBJW99}, suggesting the mechanism of collisional
deactivation. Another mechanism, involving trap excitation by a
two-step curve-crossing process, was suggested in Refs.\
\cite{AV99,MJT99}. We study below in more detail the combined
effect of both mechanisms. As the MIT experiments
\cite{IASMSK98,SIAMSK99} were conducted with Na  atoms, we shall
refer here for definiteness to Na only, though this study may be
extended to similar systems.

Given a stationary Zeeman shift, a population of molecules can
be formed as a temporary stage in the elastic process
\begin{eqnarray}
\text{Na}(\text{BEC})+\text{Na}(\text{BEC})\rightarrow
 \text{Na}_{2}\left( m\right)  \nonumber
\\
\rightarrow \text{Na(BEC)}+\text{Na}(\text{BEC}) . \label{RCol}
\end{eqnarray}
However, in the absence of other intervening interactions, or of
a time-varying field, this molecular population cannot persist, and
no loss would occur.

The first loss mechanism considered here involves the
deactivation of the resonance state, which is usually a highly
excited vibrational level in a given spin state of the pair, by an
exoergic collision with a third atom of the condensate
\cite{TTHK99,YBJW99,AV99},
\begin{equation}
\text{Na}_{2}(m)+\text{Na}(\text{BEC})\rightarrow \text{Na}_{2}(d)
+\text{Na}(\text{hot}) , \label{SCol}
\end{equation}
bringing the molecule down to a stable state $d$, and releasing
kinetic energy to the relative motion of the reaction products.
Although the collision occurs with a vanishingly small kinetic energy,
rates of such inelastic processes remain finite at near-zero energies
\cite{Forrey99}. This process naturally depends on the initial density
in the condensate. A variant of this process, involving deactivation
by collisions with another molecule (rather than an atom), of the type
\begin{equation}
\text{Na}_{2}\left( m\right) +\text{Na}_{2}\left( m\right)
 \rightarrow \text{Na}_{2}\left( d\right) +\text{Na}_{2}\left(
 u\right)  . \label{MCol}
\end{equation}
would require a significant molecular density. The two molecules
emerge in two states $d$ and $u$, where $u$ can be a stable molecular
 state
above $d$, or a continuum state of a dissociating molecule. The
 reaction
can take place as long as the corresponding internal energies obey the
inequality $E_{d}+E_{u}\le 0$, where the internal energy of $m$
 serves as the zero
reference point on the energy scale. A particularly effective reaction
of type (\ref{MCol}) would occur in the near-resonant case, in which
$0<E_{u}<|E_{d}|$. A typical example, common in VV-relaxation, is
 that of
$v+v\rightarrow \left( v+1\right) +\left( v-1\right) $ (where $v$ is
 the vibrational quantum number of the
state $m$). In this example, the kinetic energy is provided by the
vibrational anharmonicity.

Both reactions (\ref{SCol}) and (\ref{MCol}) are thus exoergic,
providing products with sufficient kinetic energy to escape the trap,
 as the
characteristic transition energies exceed the trap depth. The kinetic
 energy
may even be sufficient to produce an additional loss mechanism ---
 secondary
collisions of the reaction products with condensate atoms (see Ref.\
\cite{YBJW99}). The loss mechanism described here can be enhanced by
 bringing
close to each other the energies of the two states involved in
 reaction
(\ref{RCol}) --- the BEC state of an atom pair and the resonant
 molecular
state $m$ ---  by the application of a time-varying Zeeman shift. It
 is not
necessary that the energies of these states should cross. An actual
 crossing
of the two states can cause an irreversible transfer of population
 from the
condensate to the molecular states (see Refs.\
 \cite{YBJW99,AV99,MJT99}).
This crossing is also necessary, as the first step, in the other
 mechanism
referred to earlier --- that of excitation of the trap states by a
 two-step
crossing. The first crossing between the condensate and molecular
 states can
occur in two directions, either by letting the molecular state move
 upwards
in its energy, with respect to the condensate atom-pair state, or by
 letting
it move downwards. In both cases the loss of the molecules can
 proceed via
the deactivation mechanism. But only the upwards move alone can
 initiate the
excitation mechanism. The second crossing, at a higher energy, then
 causes
transfer of population  to higher atomic states --- bound and
 continuum trap
states in a condensate imbedded in an optical trap \cite{MJT99}, or
 continuum
states in a free condensate \cite{AV99}. This process is accompanied
 by an
increase in energy
\begin{equation}
\text{Na}(\text{BEC})+\text{Na}(\text{BEC})\rightarrow
 \text{Na}_{2}\left( m\right) \rightarrow \text{Na}(\text{hot})
+\text{Na}(\text{hot}) \label{CCros} .
\end{equation}
(To be more precise, two-step excitation can in principle occur
also in a downward move by the so-called ``counterintuitive'' process
in which the second crossing precedes the first one along the time
axis in a $Z$-shaped formation \cite{YBJB99}. However, this effect is
negligibly small in the present case.)

It should be made clear that both crossing directions have been
taken into account in Ref.\ \cite{MJT99}. However, that work does not
specify what happens to the molecular population in the downward move.
In principle, once the Zeeman shift undergoes the first crossing, the
formation of a molecular population in the resonant state can be
considered as a valid loss channel, no matter whether there are
deactivating collisions or not. In that case, the loss rate should be
independent of the deactivation rate. Our analysis here shows that
this is generally not the case. In the case of the fast-sweep
experiment, the loss would become independent of the deactivation rate
only under certain conditions (referred to in Sec.\ \ref{PasRes} below
as the ``asymptotic'' conditions).

It is now understood \cite{YBJW99,AV99} that in the slow-sweep
experiment the loss is produced almost exclusively by deactivating
collisions, such as the atom-molecular collisions described by
(\ref{SCol}). We study here also the added effect of molecule-molecule
collisions described by (\ref{MCol}). In the case of the fast-sweep
experiment, it has been claimed \cite{AV99,MJT99} that the main cause
of loss is due to the excitation processes of the kind described by
(\ref{CCros}), but it was shown \cite{YBJW99} that deactivating
collisions cannot be discounted as a contributing mechanism.

This paper therefore aims to study the effect of combining the
two kinds of mechanisms (condensate excitation vs. deactivating
collisions) together. One of the major conclusions (discussed in Sec.\
\ref{SecDisc} below) is that the two processes may actually compete
nonadditively with each other, rather then contribute additively to
the loss process.

The paper begins, in Sec.\ \ref{SecMod}, by an expansion of the
theoretical analysis used in \cite{YBJW99} to describe the effect of
deactivating collisions. We show, among other things, how the coupled
equations of the Gross-Pitaevskii type for the atomic and molecular
condensates, introduced and studied earlier by Timmermans {\it et al.}
\cite{TTHK99}, can be derived by an elimination of the product states
of the deactivation process (the so-called ``dump'' states). The
equations are then extended to include molecule-molecule collisions of
type (\ref{MCol}). In Sec.\ \ref{SecExSt} we add to the deactivation
model an effect representing the outcome of the excitation process.
The results are presented and discussed in Sec.\ \ref{SecDisc}, in
comparison with the MIT experiments.

\section{The deactivation model} \label{SecMod}

\subsection{Hamiltonian and variational procedure}

Let us consider an optically-trapped BEC exposed to an external
homogeneous time-dependent magnetic field $B\left( t\right) $ used to
 tune a
vibrationally excited molecular state $m$ to a Feshbach resonance with
the state of a pair of unbound condensate atoms. In order to write
down an hamiltonian for such a system, including the molecular dump
states, we must introduce field annihilation operators of the atoms
$\hat{\psi }\left( {\bf r}\right) $, of the molecular resonant state
$\hat{\psi }^{+}_{m}\left( {\bf r}_{m}\right) $, and of the lower and
 upper dump states, $\hat{\psi }^{+}_{d}\left( {\bf r}_{m}\right) $
 and $\hat{\psi }^{+}_{u}\left( {\bf r}_{m}\right) $,
respectively [see discussion following Eq.\ (\ref{MCol})]. The
hamiltonian can then be written as
\begin{eqnarray}
\hat{H}=&&\int d^{3}r\hat{\psi }^{+}\left( {\bf r}\right)
 \hat{H}_{a}\hat{\psi }\left( {\bf r}\right) +\hat{V}_{el} \nonumber
\\
&&+\int d^{3}r_{m}\sum\limits^{}_{\alpha =m,u,d}\hat{\psi }^{
+}_{\alpha }\left( {\bf r}_{m}\right) \hat{H}_{\alpha }\hat{\psi
 }_{\alpha }\left( {\bf r}_{m}\right)  \nonumber
\\
&&+\hat{V}_{h}+\hat{V}^{+}_{h}+\sum\limits^{}_{d}\left( \hat{V}_{d}
+\hat{V}^{+}_{d}\right) +\sum\limits^{}_{ud}\left( \hat{V}_{ud}
+\hat{V}^{+}_{ud}\right)  . \label{SQH}
\end{eqnarray}
The terms
\begin{eqnarray}
&&\hat{H}_{a}={1\over 2m}\hat{{\bf p}}^{2}+V_{a}\left( {\bf r}\right)
 -\mu _{a}B\left( t\right)  , \nonumber
\\
&&
\\
&&\hat{H}_{\alpha }={1\over 4m}\hat{{\bf p}}^{2}_{m}+V_{\alpha }\left
( {\bf r}_{m}\right) -\mu _{\alpha }B\left( t\right)  , \nonumber
\end{eqnarray}
(where $\alpha =m, u$, or $d$, includes the resonant state) are the
hamiltonians for the noninteracting atoms and molecules. Here
 $V_{a}\left( {\bf r}\right) $
and $V_{\alpha }\left( {\bf r}_{m}\right) $ are the corresponding
 atomic and molecular energies as
functions of the position ${\bf r}$ of the atomic (or ${\bf r}_{m}$
 of the molecular)
center of mass, whose values include the optical trap potentials and
the position-independent differences of internal energies in the
absence of the trap. Also, $\mu _{a}$and $\mu _{\alpha }$  are the
 corresponding magnetic
moments.

Since the atoms and molecules are treated here as independent
particles, the interaction responsible for the atom-molecule coupling
[reaction (\ref{RCol})] can be written in the general form
\begin{mathletters}
\begin{equation}
\hat{V}_{h}=\int d^{3}r d^{3}r^\prime  V_{h}\left( {\bf r}-{\bf
 r}^\prime \right)  \hat{\psi }^{+}_{m}\left( {{\bf r}+{\bf r}^\prime
 \over 2}\right) \hat{\psi }\left( {\bf r}\right) \hat{\psi }\left(
 {\bf r}^\prime \right)  , \label{Vh}
\end{equation}
in which the molecule, created as an independent particle
preserves the position of the center of mass. However, considering
that the interaction is localized within a range of atomic size,
negligibly small compared to the condensate size and the relevant de
Broglie wavelengths, one can use the approximation of zero-range
interaction $V_{h}\left( {\bf r}-{\bf r}^\prime \right) =g\delta
 \left( {\bf r}-{\bf r}^\prime \right) $, and represent the
 interaction in the
simpler form
\begin{equation}
\hat{V}_{h}=g\int d^{3}r \hat{\psi }^{+}_{m}\left( {\bf r}\right)
 \hat{\psi }\left( {\bf r}\right) \hat{\psi }\left( {\bf r}\right)  .
 \label{VH}
\end{equation}
\end{mathletters} The same arguments are applicable to the terms
in Eq.\ (\ref{SQH}) representing the deactivating collisions
 $\hat{V}_{d}$  and
$\hat{V}_{ud}$  [reactions (\ref{SCol}) and (\ref{MCol}),
 respectively].
However, the use of a zero-range interaction would lead to a
divergence in the ensuing calculations [see discussion following Eq.\
(\ref{phidkro}) below]. Therefore we keep these interactions as
finite-range functions of the distance between the reaction products,
writing
\begin{eqnarray}
\hat{V}_{d}=\int d^{3}rd^{3}r_{m}&&d_{d}\left( |{\bf r}-{\bf r}_{m}
|\right) \hat{\psi }^{+}\left( {\bf r}\right) \hat{\psi }^{
+}_{d}\left( {\bf r}_{m}\right)  \nonumber
\\
&&\times \hat{\psi }\left( {{\bf r}+2{\bf r}{ } _{m}\over 3}\right)
 \hat{\psi }_{m}\left( {{\bf r}+2{\bf r}{ } _{m}\over 3}\right)
 \label{Vd}
\\
\hat{V}_{ud}=\int d^{3}r_{1}d^{3}r_{2}&&d_{ud}\left( |{\bf r}_{1}
-{\bf r}_{2}|\right) \hat{\psi }^{+}_{u}\left( {\bf r}_{1}\right)
 \hat{\psi }^{+}_{d}\left( {\bf r}_{2}\right)  \nonumber
\\
&&\times \hat{\psi }_{m}\left( {{\bf r}_{1}+{\bf r}{ } _{2}\over
 2}\right) \hat{\psi }_{m}\left( {{\bf r}_{1}+{\bf r}{ } _{2}\over
 2}\right)  . \label{Vud}
\end{eqnarray}
Here, as in (\ref{Vh}) the position of the center of mass is
preserved, but the finite-range nature of the interactions is retained
in the functions $d_{d}\left( \rho \right) $ and $d_{ud}\left( \rho
 \right) $, whose actual shape will be
discussed further down.

Finally, the part of the hamiltonian associated with elastic
collisions (see Ref.\ \cite{TTHK99}),
\begin{eqnarray}
\hat{V}_{el}=\int d^{3}r \biggl\lbrack && {U{ } _{a}\over 2}\hat{\psi
 }^{+}\left( {\bf r}\right) \hat{\psi }^{+}\left( {\bf r}\right)
 \hat{\psi }\left( {\bf r}\right) \hat{\psi }\left( {\bf r}\right)
 \nonumber
\\
&&+{U{ } _{m}\over 2}\sum\limits^{}_{\alpha ,\alpha ^\prime
 }\hat{\psi }^{+}_{\alpha }\left( {\bf r}\right) \hat{\psi }^{
+}_{\alpha ^\prime }\left( {\bf r}\right) \hat{\psi }_{\alpha ^\prime
 }\left( {\bf r}\right) \hat{\psi }_{\alpha }\left( {\bf r}\right)
 \nonumber
\\
&&+U_{am}\sum\limits^{}_{\alpha }\hat{\psi }^{+}\left( {\bf
 r}\right) \hat{\psi }^{+}_{\alpha }\left( {\bf r}\right) \hat{\psi
 }_{\alpha }\left( {\bf r}\right) \hat{\psi }\left( {\bf r}\right)
 \biggr\rbrack  ,
\end{eqnarray}
includes terms proportional to the zero-momentum atom-atom,
molecule-molecule, and atom-molecule interactions,
\begin{equation}
U_{a}={4\pi \hbar { } ^{2}\over m}a_{a},\quad U_{m}={2\pi \hbar { }
 ^{2}\over m}a_{m},\quad U_{am}={3\pi \hbar { } ^{2}\over
 m}a_{am} ,
\end{equation}
where $a_{a}$, $a_{m}$, and $a_{am}$  are the corresponding
 elastic scattering
lengths, and the different numerical factors in the numerators reflect
the different reduced masses.

We outline here the derivation of mean-field equations, involving
$c$-number fields, that represent all the actively participating
 states
listed above. This is accomplished by an extension of a well-known
variational method (see Refs.\ \cite{BR86,EDCB96}). Let us introduce
the trial wavefunction
\begin{eqnarray}
|\Phi \rangle =&&\Bigl\lbrack 1+\int d^{3}r
 d^{3}r_{m}\sum\limits^{}_{d}\varphi _{d}\left( {\bf r},{\bf
 r}_{m},t\right) \hat{\psi }^{+}\left( {\bf r}\right) \hat{\psi }^{
+}_{d}\left( {\bf r}_{m}\right)  \nonumber
\\
&&+\int d^{3}r_{1}d^{3}r_{2}\sum\limits^{}_{ud}\varphi _{ud}\left(
 {\bf r}_{1},{\bf r}_{2},t\right) \hat{\psi }^{+}_{u}\left( {\bf
 r}_{1}\right) \hat{\psi }^{+}_{d}\left( {\bf r}_{2}\right)
 \Bigr\rbrack  \nonumber
\\
\times &&|\varphi _{0},\varphi _{m}\rangle  . \label{TrFun}
\end{eqnarray}
The factor
\begin{eqnarray}
|\varphi _{0},\varphi _{m}\rangle =\exp\Bigl\{ \int
 d^{3}r\bigl\lbrack &&\varphi _{0}\left( {\bf r},t\right) \hat{\psi
 }^{+}\left( {\bf r}\right)  \nonumber
\\
&&+\varphi _{m}\left( {\bf r},t\right) \hat{\psi }^{+}_{m}\left( {\bf
 r}\right) \bigr\rbrack \Bigr\}|0\rangle
\end{eqnarray}
\onecolumn
\widetext
\noindent is a coherent state, formed by a product of exponential
operators, involving atomic ($\varphi _{0}$) and molecular ($\varphi
 _{m}$) condensate states,
and operating on the vacuum state $|0\rangle $. The linear factor
 preceding it
in Eq.\ (\ref{TrFun}) includes the fields $\varphi _{d}\left( {\bf
 r},{\bf r}_{2},t\right) $ and $\varphi _{ud}\left( {\bf r}_{1},{\bf
 r}_{2},t\right) $
which are the correlated states of the products in reactions
(\ref{SCol}) and (\ref{MCol}), respectively.

The linear form of the latter factor forces the trial
wavefunction (\ref{TrFun}) to take into account only one-particle
occupation of the non-resonant molecular states $u$ and $d$, as a
constraint. This approximation is based on the assumption of fast
removal of ``hot'' particles from the trap. In contrast, the
occupation of the resonant state ($m$) is allowed to reach the order
 of
magnitude of the condensate-state occupation (as our calculations
verify). Therefore $\varphi _{m}$  describes a coherent molecular
 condensate.

Another constraint, that follows from the large energy difference
between the dump states and the resonant state, is the condition
\begin{equation}
\int d^{3}r\varphi ^{*}_{0}\left( {\bf r},t\right) \varphi _{d}\left(
 {\bf r},{\bf r}_{m},t\right) =0 \label{ortphi}
\end{equation}
[see discussion following Eq.\ (\ref{phidkro}) for
justification]. The trial wavefunction (\ref{TrFun}) can now be
substituted into a variational functional (see Refs.\
\cite{BR86,EDCB96})
\begin{equation}
\int\limits^{\infty }_{-\infty }dt{\langle \Phi |i\hbar  {\partial
 \over \partial t}-\hat{H}|\Phi \rangle \over \langle \Phi |\Phi
 \rangle } .
\end{equation}
Neglecting terms of the order of $\varphi ^{3}_{d}$  and $\varphi
 ^{3}_{u}$, and taking
(\ref{ortphi}) into account, the use of a standard variational
procedure (see Ref.\ \cite{EDCB96}) then leads to a set of coupled
equations for the atomic ($\varphi _{0}$) and molecular ($\varphi
 _{m}$) condensate fields (or
``wavefunctions''), as well as for the dump states ($\varphi _{d}$
 and $\varphi _{ud}$):
%\twocolumn[\widetext 
\begin{mathletters}
\begin{eqnarray}
i\hbar \dot{\varphi }_{0}\left( {\bf r},t\right) =&&\left(
 \hat{H}_{a}+U_{a}|\varphi _{0}\left( {\bf r},t\right) |^{2}+U_{am}
|\varphi _{m}\left( {\bf r},t\right) |^{2}\right) \varphi _{0}\left(
 {\bf r},t\right) +2g^{*}\varphi ^{*}_{0}\left( {\bf r},t\right)
 \varphi _{m}\left( {\bf r},t\right) +Q\left( {\bf r},t\right)
 \varphi ^{*}_{m}\left( {\bf r},t\right)  \label{phi0d}
\\
i\hbar \dot{\varphi }_{m}\left( {\bf r},t\right) =&&\left(
 \hat{H}_{m}+U_{am}|\varphi _{0}\left( {\bf r},t\right) |^{2}+U_{m}
|\varphi _{m}\left( {\bf r},t\right) |^{2}\right) \varphi _{m}\left(
 {\bf r},t\right) +g\varphi ^{2}_{0}\left( {\bf r},t\right)  \nonumber
\\
&&+Q\left( {\bf r},t\right) \varphi ^{*}_{0}\left( {\bf r},t\right)
+Q_{m}\left( {\bf r},t\right) \varphi ^{*}_{m}\left( {\bf r},t\right)
  \label{phimd}
\\
i\hbar \dot{\varphi }_{d}\left( {\bf r}_{1},{\bf r}_{2},t\right)
 =&&\left( {1\over 2m}\hat{p}^{2}_{1}+{1\over 4m}\hat{p}^{2}_{2}
-E_{d}\right) \varphi _{d}\left( {\bf r}_{1},{\bf r}_{2},t\right)
+d_{d}\left( |{\bf r}_{1}-{\bf r}_{2}|\right) \varphi _{0}\left(
 {{\bf r}_{1}+2{\bf r}{ } _{2}\over 3},t\right) \varphi _{m}\left(
 {{\bf r}_{1}+2{\bf r}{ } _{2}\over 3},t\right)  \label{phid}
\\
i\hbar \dot{\varphi }_{ud}\left( {\bf r}_{1},{\bf r}_{2},t\right)
 =&&\left( {1\over 4m}\hat{p}^{2}_{1}+{1\over 4m}\hat{p}^{2}_{2}
-E_{u}-E_{d}\right) \varphi _{ud}\left( {\bf r}_{1},{\bf
 r}_{2},t\right) +d_{ud}\left( |{\bf r}_{1}-{\bf r}_{2}|\right)
 \varphi ^{2}_{m}\left( {{\bf r}_{1}+{\bf r}{ } _{2}\over 2},t\right)
  , \label{phiud}
\end{eqnarray}
\end{mathletters} where
\begin{eqnarray}
&&Q\left( {\bf r},t\right) =\int
 d^{3}r_{1}d^{3}r_{2}\sum\limits^{}_{d}d^{*}_{d}\left( |{\bf r}_{1}
-{\bf r}_{2}|\right) \varphi _{d}\left( {\bf r}_{1},{\bf
 r}_{2},t\right) \delta \left( {\bf r}-{{\bf r}_{1}+2{\bf r}{ }
 _{2}\over 3}\right)  \label{Q}
\\
&&Q_{m}\left( {\bf r},t\right) =2\int
 d^{3}r_{1}d^{3}r_{2}\sum\limits^{}_{ud}d^{*}_{ud}\left( |{\bf r}_{1}
-{\bf r}_{2}|\right) \varphi ^{*}_{m}\left( {\bf r},t\right) \varphi
 _{ud}\left( {\bf r}_{1},{\bf r}_{2},t\right) \delta \left( {\bf r}
-{{\bf r}_{1}+{\bf r}{ } _{2}\over 2}\right)  . \label{Qm}
\end{eqnarray}
The terms corresponding to elastic collisions, dependent on $|\varphi
 _{d}|^{2}$
and $|\varphi _{ud}|^{2}$, which should appear in Eqs.\ (\ref{phi0d})
 and
(\ref{phimd}), are omitted since they are negligible compared to the
terms including $\varphi _{0}$  and $\varphi _{m}$. In Eqs.\
(\ref{phid}) and (\ref{phiud})
the terms corresponding to elastic collisions, Zeeman shifts, and trap
potentials are neglected compared to the position-independent internal
energies, denoted here $E_{d}$  and $E_{u}$.

\subsection{Dump state elimination}

The procedure used to eliminate the dump states is similar to the
Weisskopf-Wigner method of the theory of spontaneous emission (see
Ref.\ \cite{Agarwal}). Equation (\ref{phid}) is of the form of a
Schr\"odinger equation for two free particles with a source (the last
term in the right-hand side). Such an equation can be solved by
applying the Green's function method for free particles, with the
result
\begin{eqnarray}
\varphi _{d}\left( {\bf r}_{1},{\bf r}_{2},t\right) =&&-{i\over \left
( 2\pi \right) ^{6}\hbar }\int\limits^{t}_{-\infty }dt^\prime \int
 d^{3}K d^{3}k \exp\left\lbrack -i\left( {\hbar K{ } ^{2}\over 6m}
+{3\hbar k{ } ^{2}\over 4m}-{E{ } _{d}\over \hbar }-i0\right) \left(
 t-t^\prime \right) \right\rbrack  \nonumber
\\
&&\times \exp\left\lbrack i{\bf K}{{\bf r}_{1}+2{\bf r}{ } _{2}\over
 3}+i{\bf k}\left( {\bf r}_{1}-{\bf r}_{2}\right) \right\rbrack  \int
 d^{3}\rho  d_{d}\left( \rho \right) e^{-i{\bf k}\bbox{\rho}}\int
 d^{3}R e^{-i{\bf K}{\bf R}}\varphi _{m}\left( {\bf R},t^\prime
 \right) \varphi _{0}\left( {\bf R},t^\prime \right)  . \label{phidGf}
\end{eqnarray}
\twocolumn[\widetext
Here ${\bf R}$ is the center-of-mass position of the three-atom
 system,
$\bbox{\rho}$ is the radius vector of the reaction products, and
 ${\bf K}$, ${\bf k}$
are the corresponding wavevectors. Since $\varphi _{m}\left( {\bf
 R},t\right) $ and $\varphi _{0}\left( {\bf R},t\right) $ are
condensate wavefunctions, the Fourier transform of their product
\lbrack the integral over ${\bf R}$ in Eq.\ (\ref{phidGf})\rbrack
 vanishes if
$K>1/b$, where $b$ is a characteristic size of the condensate.
 Therefore
$\hbar ^{2}K^{2}/\left( 6m\right) <\hbar \omega _{\text{trap}}$  is
 negligible compared to $E_{d}\gg \hbar \omega _{\text{trap}}$, where
 $\omega _{\text{trap}}$  is
the trap frequency. This fact allows us also to neglect the time
dependence of $\varphi _{m}\left( {\bf R},t\right) $ and $\varphi
 _{0}\left( {\bf R},t\right) $ in the integration over $t^\prime $,
 and
thus obtain the simplified expression
\begin{equation}
\varphi _{d}\left( {\bf r}_{1},{\bf r}_{2},t\right) =-{1\over \left(
 2\pi \right) { } ^{3}}\varphi _{0}\left( {{\bf r}_{1}+2{\bf r}{ }
 _{2}\over 3},t\right) \varphi _{m}\left( {{\bf r}_{1}+2{\bf r}{ }
 _{2}\over 3},t\right) \int d^{3}k d^{3}\rho  {d_{d}\left( \rho
 \right) \exp\left\lbrack i{\bf k}\left( {\bf r}_{2}-{\bf r}_{1}
-\bbox{\rho}\right) \right\rbrack \over 3\hbar ^{2}k^{2}/\left(
 4m\right) -E_{d}-i0} . \label{phidkro}
\end{equation}
] \narrowtext The atom-molecule pair is thus formed with a
momentum $\hbar k_{d}=\sqrt{4mE_{d}/3}$ of relative motion. The
 function $\varphi _{d}\left( {\bf r}_{1},{\bf r}_{2},t\right) $ is a
rapidly oscillating function of the coordinates, and therefore
condition (\ref{ortphi}) is justified.

Substituting Eq.\ (\ref{phidkro}) into Eqs.\ (\ref{Q}), and
(\ref{Qm}) and introducing a Fourier transform of the function
 $d_{d}\left( \rho \right) $
\begin{equation}
\tilde{d}_{d}\left( k\right) =\int d^{3}\rho d_{d}\left( \rho \right)
 \exp\left( -i{\bf k}\bbox{\rho}\right)  ,
\end{equation}
one obtains
\begin{equation}
Q\left( {\bf r},t\right) =-\left( \delta +i\hbar \gamma \right)
 \varphi _{0}\left( {\bf r},t\right) \varphi _{m}\left( {\bf
 r},t\right)  , \label{QDelGam}
\end{equation}
where
\begin{equation}
\delta +i\hbar \gamma ={1\over \left( 2\pi \right) { }
 ^{3}}\sum\limits^{}_{d} \int d^{3}k {|\tilde{d}_{d}\left( k\right)
|{ } ^{2}\over 3\hbar ^{2}k^{2}/\left( 4m\right) -E_{d}-i0} .
\end{equation}
Using the well-known identity $\left( x-i0\right) ^{-1}={\cal P}x^{
-1}+i\pi \delta \left( x\right) $, where ${\cal P}$
denotes the Cauchy principal part of the integral, allows us to obtain
explicit expressions for $\gamma $ and $\delta $:
\begin{eqnarray}
\gamma &&={m\over 3\pi \hbar { } ^{3}} \sum\limits^{}_{d}k_{d}\left
|\tilde{d}_{d}\left( k_{d}\right) \right|^{2} \label{gamma}
\\
\delta &&={2m\over 3\pi ^{2}\hbar { } ^{2}}\sum\limits^{}_{d}{\cal
 P}\int\limits^{\infty }_{0}dk{k^{2}|\tilde{d}_{d}\left( k\right) |{
 } ^{2}\over k^{2}-k{ } ^{2}_{d}} . \label{delta}
\end{eqnarray}
A similar analysis, starting from Eq.\ (\ref{phiud}), gives
\begin{equation}
Q_{m}\left( {\bf r},t\right) =-\left( \delta _{m}+i\hbar \gamma
 _{m}\right) \varphi ^{2}_{m}\left( {\bf r},t\right)  ,
 \label{QmDelGam}
\end{equation}
where
\begin{eqnarray}
\gamma _{m}=&&{m\over \pi \hbar { }
 ^{3}}\sum\limits^{}_{u,d}k_{ud}\left|\tilde{d}_{ud}\left(
 k_{ud}\right) \right|^{2} \label{gammam}
\\
\delta _{m}=&&{2m\over \pi ^{2}\hbar { }
 ^{2}}\sum\limits^{}_{u,d}{\cal P}\int\limits^{\infty }_{0}dk{k^{2}
|\tilde{d}_{ud}\left( k\right) |{ } ^{2}\over k^{2}-k{ } ^{2}_{ud}} .
 \label{deltam}
\end{eqnarray}
and $k_{ud}=\sqrt{2m\left( E_{u}+E_{d}\right) }/\hbar $.
 Substituting Eqs.\ (\ref{QDelGam}) and
(\ref{QmDelGam}) into Eqs.\ (\ref{phi0d}) and (\ref{phimd}) one
finally obtains a pair of coupled Gross-Pitaevskii equations (see
Ref.\ \cite{GP}) \begin{mathletters} \label{GPE}
\begin{eqnarray}
i\hbar \dot{\varphi }_{0}&&=\left( \hat{H}_{a}+U_{a}|\varphi _{0}
|^{2}+U_{am}|\varphi _{m}|^{2}\right) \varphi _{0}+2g^{*}\varphi ^{
*}_{0}\varphi _{m} \nonumber
\\
&&-\left( \delta +i\hbar \gamma \right) |\varphi _{m}|^{2}\varphi
 _{0} \label{GPEa}
\\
i\hbar \dot{\varphi }_{m}&&=\left( \hat{H}_{m}+U_{am}|\varphi _{0}
|^{2}+U_{m}|\varphi _{m}|^{2}\right) \varphi _{m}+g\varphi ^{2}_{0}
 \nonumber
\\
&&-\left\lbrack \left( \delta +i\hbar \gamma \right) |\varphi _{0}
|^{2}+\left( \delta _{m}+i\hbar \gamma _{m}\right) |\varphi _{m}
|^{2}\right\rbrack \varphi _{m} . \label{GPEm}
\end{eqnarray}
\end{mathletters} The parameters $\delta $, $\gamma $, $\delta _{m}$,
 and $\gamma _{m}$, which are
expressed in terms of $d_{d}$  and $d_{ud}$  [see Eqs.\ (\ref{gamma}),
(\ref{delta}), (\ref{gammam}), and (\ref{deltam})], describe the shift
and the width of the resonance due to the deactivating collisions with
atoms and molecules, respectively. The parameters $\gamma $ and
 $\gamma _{m}$  are one
half of the corresponding rate constants. Since the strengths of the
deactivating interactions are unknown, the parameter $\gamma $ will be
extracted below from the experimental data, and $\gamma _{m}$  will
 be used below
as an adjustable parameter. The shifts $\delta $ and $\delta _{m}$
 can be incorporated
in the interactions $U_{am}$  and $U_{m}$, respectively.

Equations (\ref{GPE}) are similar to those presented recently
by Timmermans {\it et al.} \cite{TTHK99}. Among other things, Ref.\
\cite{TTHK99} shows that in the case of a time-independent magnetic
field and large resonant detuning, neglecting the decay described
by the imaginary terms, Eqs.\ (\ref{GPE}) can be reduced to a
single Gross-Pitaevskii equation with an effective scattering
length $a_{a}\left\lbrack 1-\Delta /\left( B-B_{0}\right)
 \right\rbrack $, where the parameter $\Delta $ is related to the
atom-molecule coupling constant $g$ of Eq.\ (\ref{VH}) as
\begin{equation}
|g|^{2}=2\pi \hbar ^{2}|a_{a}|\mu \Delta /m . \label{Delta}
\end{equation}
Values of $\Delta $ for Na were calculated in Refs.\
 \cite{MJT99,AV99},
or extracted from the experimental data in Refs.\
\cite{IASMSK98,SIAMSK99}.

\subsection{Density equations and approximate solutions}

Let us introduce the new real variables
\begin{eqnarray}
&&n\left( {\bf r},t\right) =|\varphi _{0}\left( {\bf r},t\right)
|^{2}, \quad n_{m}\left( {\bf r},t\right) =|\varphi _{m}\left( {\bf
 r},t\right) |^{2}, \nonumber
\\
&&u\left( {\bf r},t\right) =2\text{Re}\left( g\varphi ^{2}_{0}\left(
 {\bf r},t\right) \varphi ^{*}_{m}\left( {\bf r},t\right) \right)
/\hbar  ,
\\
&&v\left( {\bf r},t\right) =-2\text{Im}\left( g\varphi ^{2}_{0}\left(
 {\bf r},t\right) \varphi ^{*}_{m}\left( {\bf r},t\right) \right)
/\hbar  . \nonumber
\end{eqnarray}
The time evolution of these variables can be described by a set
of real equations, similar to the optical Bloch equations where $n$
 and
$n_{m}$  act like ``populations'' and $u$ and $v$ like
 ``coherences''. When
the kinetic energy terms are  neglected, in accordance with the
Thomas-Fermi approximation (see Refs.\ \cite{PW98,DGPS99}), one
obtains from Eqs.\ (\ref{GPE}) \begin{mathletters} \label{deneq}
\begin{eqnarray}
&&\dot{n}=2v-2\Gamma _{a}n \label{deneqn}
\\
&&\dot{n}_{m}=-v-2\Gamma _{m}n_{m}
\\
&&\dot{v}=Du-\left( 2\Gamma _{a}+\Gamma _{m}\right) v+2|g|^{2}n\left(
 4n_{m}-n\right) /\hbar ^{2}
\\
&&\dot{u}=-Dv-\left( 2\Gamma _{a}+\Gamma _{m}\right) u .
\end{eqnarray}
\end{mathletters} Here
\begin{eqnarray}
&&D\left( {\bf r},t\right) =\{V\left( {\bf r}\right) -\mu B\left(
 t\right)  \nonumber
\\
&&+2\left\lbrack U_{a}n\left( {\bf r},t\right) +\left( U_{am}
-\delta \right) n_{m}\left( {\bf r},t\right) \right\rbrack  \nonumber
\\
&&-\left\lbrack \left( U_{am}-\delta \right) n\left( {\bf
 r},t\right) +\left( U_{m}-\delta _{m}\right) n_{m}\left( {\bf
 r},t\right) \right\rbrack \}/\hbar  , \label{D}
\\
&&V\left( {\bf r}\right) =2V_{a}\left( {\bf r}\right) -V_{m}\left(
 {\bf r}\right) , \quad \mu =2\mu _{a}-\mu _{m}, \nonumber
\end{eqnarray}
and
\begin{eqnarray}
&&\Gamma _{a}\left( {\bf r},t\right) =\gamma n_{m}\left( {\bf
 r},t\right)  , \nonumber
\\
&&
\\
&&\Gamma _{m}\left( {\bf r},t\right) =\gamma n\left( {\bf r},t\right)
 +\gamma _{m}n_{m}\left( {\bf r},t\right)  . \nonumber
\end{eqnarray}
In the Thomas-Fermi approximation the functions $n\left( {\bf
 r},t\right) $, $n_{m}\left( {\bf r},t\right) $,
$v\left( {\bf r},t\right) $, and $u\left( {\bf r},t\right) $ depend
 on ${\bf r}$ only parametrically. The set of four
real equations (\ref{deneq}) can then be solved numerically using as
initial conditions either an ${\bf r}$-dependent (for example, a
 steady-state
Thomas-Fermi) distribution, or an ${\bf r}$-independent (homogeneous)
distribution equal to the mean trap density.

Nevertheless, certain properties of the solutions can be derived
analytically from Eqs.\ (\ref{deneq}), without recourse to numerical
solutions, whenever the following ``fast decay'' conditions hold:
\begin{mathletters} \label{FDA}
\begin{eqnarray}
&&\mu \dot{B}\ll \hbar \Gamma _{m}\left( D+\Gamma ^{2}_{m}/D\right)
 ,\quad \Gamma _{m}\gg \Gamma _{a}
\\
&&D^{2}+\Gamma ^{2}_{m}\gg 6|g|^{2}n/\hbar ^{2} . \label{FDAb}
\end{eqnarray}
\end{mathletters} These conditions mean that the relaxation of
$n_{m}$, $v$, and $u$ is much faster compared to that of $n$ and to
 the rate of
change of the energy, caused by the magnetic field with a sweep rate
$\dot{B}$. Therefore the values of the fast variables can be related
 to a
given $n$ value, using a quasi-stationary approximation, by
\begin{eqnarray}
&&u\sim -{D\over \Gamma { } _{m}}v ,\qquad v\sim -{2|g
|^{2}n^{2}\Gamma { } _{m}\over \hbar ^{2}\left( D^{2}+\Gamma
 ^{2}_{m}\right) } , \nonumber
\\
&&\label{quasist}
\\
&&n_{m}\sim {|g|^{2}n{ } ^{2}\over \hbar ^{2}\left( D^{2}+\Gamma
 ^{2}_{m}\right) } , \nonumber
\end{eqnarray}
and the condition (\ref{FDAb}) leads to $n_{m}\ll n$. As a result, a
single non-linear rate equation for the atomic density can be
extracted. When terms proportional to the atomic and molecular
densities in $D$ [see Eq.\ (\ref{D})] are neglected, the resulting
 rate
equation is
\begin{equation}
\dot{n}\left( {\bf r},t\right) =-{6|g|^{2}\gamma n^{3}\left( {\bf
 r},t\right) \over \left\lbrack V\left( {\bf r}\right) -\mu B\left(
 t\right) \right\rbrack ^{2}+\left\lbrack \hbar \gamma n\left( {\bf
 r},t\right) \right\rbrack { } ^{2}} . \label{RE}
\end{equation}
(The neglected terms in $D$ effectively add an extra shift to the
resonance, but its contribution is hardly noticed in the present
problem.)

Equation (\ref{RE}) has a form analogous to the Breit-Wigner
expression for resonant scattering in the limit of zero-momentum
collisions (see Ref.\ \cite{MJT99}).  In the Breit-Wigner sense one
can interpret $2\hbar \gamma n$  as the width of the decay channel,
 while the width
of the input channel is proportional to $|g|^{2}$. This observation
establishes a link between the macroscopic approach used here and
microscopic approaches that treat the loss rate as a collision
process. However, Eq.\ (\ref{RE}) differs from the usual Breit-Wigner
expression by a factor ${3\over 2}$, associated with the loss of a
 third
condensate atom in the reaction (\ref{SCol}).

\subsubsection{Approaching the resonance}

Very close to resonance (e.g., for conditions prevailing in the
experiment \cite{SIAMSK99}, where $B\left( t\right) $ is within $1
 \mu $T of resonance)
the behavior of Eq.\ (\ref{RE}) effectively attains a 1-body form
linear in $n$. But as long as we stay out of this narrow region, by
obeying the ``{\it off-resonance}'' condition (which holds in the
 slow-sweep
experiment \cite{IASMSK98,SIAMSK99}),
\begin{equation}
\hbar \gamma n\left( {\bf r},t\right) \ll |V\left( {\bf r}\right)
-\mu B\left( t\right) | , \label{NRA}
\end{equation}
\noindent we can write Eq.\ (\ref{RE}) (to a very good approximation)
 in the
3-body form $\dot{n}=-K_{3}\left( {\bf r},t\right) n^{3}$, where
\begin{equation}
K_{3}={12\pi \hbar ^{2}|a_{a}|\gamma \Delta \over m\mu \left\lbrack
 B\left( t\right) -V\left( {\bf r}\right) /\mu \right\rbrack { }
 ^{2}} . \label{K3}
\end{equation}
The dependence of Eq.\ (\ref{K3}) on the scattering length $a_{a}$
follows from Eq.\ (\ref{Delta}). A similar expression has been
obtained in Ref.\ \cite{TTHK99}, but the loss of the third condensate
atom in the deactivating reaction (\ref{SCol}), described by the term
proportional to $\Gamma _{a}$  in Eq.\ (\ref{deneqn}), was neglected.
 In the
fast-decay approximation, in which $\dot{n}_{m}$  is diminishingly
 small, this
neglected term is equal to $-v$. Added to the $-2v$, this makes
 $K_{3}$  of
Eq.\ (\ref{K3}) larger by a factor ${3\over 2}$ than the
 corresponding expression
in Ref.\ \cite{TTHK99}. This omission has been corrected in Refs.\
\cite{YBJW99,AV99}.

When the magnetic field ramp is assumed to vary linearly in time,
starting at $t_{0}$  and ending at $t$, and Eq. (\ref{NRA}) applies
throughout the ramp motion (i.e., by avoiding passage through the
resonance), the rate equation can be solved analytically. Using  Eq.\
(\ref{K3}) one then gets
\begin{eqnarray}
n\left( {\bf r},t\right) =n\left( {\bf r},t_{0}\right) \biggl\lbrack
 1+24\pi \hbar ^{2}|a_{a}|\Delta \gamma n^{2}\left( {\bf
 r},t_{0}\right)  \nonumber
\\
\times \left( t-t_{0}\right) /\left( m\mu \dot{B}^{2}t t_{0}\right)
 \biggr\rbrack ^{-1/2}, \label{nt}
\end{eqnarray}
\noindent where $\dot{B}$ is the magnetic-field ramp speed and the
 extrapolated time of
reaching exact resonance is chosen to be $t=0$, so that both $t$ and
 $t_{0}$
have the same sign. We shall refer to the combination of Eqs.\
(\ref{FDA}) and (\ref{NRA}), that leads to conditions (\ref{K3}), as
the ``three-body'' approximation.

\subsubsection{Passing through the resonance} \label{PasRes}

The three-body approximation of Eqs.\ (\ref{NRA}) and (\ref{K3})
does not hold very close to resonance, and is therefore inapplicable
to the description of the fast-sweep experiment, in which the Zeeman
shift is swept rapidly {\it through} the resonance, causing dramatic
 losses
(see Refs.\ \cite{IASMSK98,SIAMSK99}). Nevertheless, the fast decay
approximation (\ref{FDA}) may still be valid. A simple analytical
expression can then be derived for the condensate loss on passage
though the resonance if, in addition, the magnetic field variation
lasts long enough to reach the ``{\it asymptotic}'' condition
\begin{equation}
\mu \delta B\gg \hbar \gamma n , \label{ascond}
\end{equation}
where $\delta B$ is the total change in $B$ accumulated over the
 sweep.
This condition allows the extension of the ramp starting and stopping
times to $\mp \infty $.

The asymptotic behavior of $n\left( t\right) $ and $\dot{n}\left(
 t\right) $  in the complex $t$ plane,
as $|t|\rightarrow \infty $, is constrained by consistency
 requirements. Consider the
four possible asymptotic relations between $n$ and $t$ shown in the
 first
column of Table \ref{tabas}. Equation (\ref{RE}) forces $\dot{n}$  to
 attain
the form in the second column, from which it follows that $n$ should
attain the form in the third column. Obviously, cases (c) (for
$\text{Re} t>0$) and (d) (at all $t$ values) are not self-consistent.
Therefore, the asymptotic solution may attain only one of the forms
complying with cases (a), (b), or (c) (the latter for $\text{Re} t<0$
only).

One can now evaluate the variation of $n\left( {\bf r},t\right) $ in
 the infinite time
interval $\left( -\infty ,\infty \right) $. Let us rewrite Eq.\
(\ref{RE}) in the form
\begin{equation}
{dn\over n{ } ^{2}}=-{6\gamma n|g|{ } ^{2}\over \left( \mu
 \dot{B}t\right) ^{2}+\left( \hbar \gamma n\right) { } ^{2}}dt ,
\end{equation}
where $V\left( {\bf r}\right) $ is removed by our choice of the
 origin on the time
scale. We then integrate the left-hand side with respect to $n$ from
$n\left( {\bf r},-\infty \right) $ to $n\left( {\bf r},\infty \right)
 $ and the right-hand side with respect to $t$ from $-\infty $ to
$\infty $, considering $n$ as a well-defined function of $t$. The
 latter integral
may be evaluated by using the residue theorem, closing the integration
contour by an arc of infinite radius in the upper half-plane. The
integral along this arc vanishes according to the asymptotic behavior
of $n$ considered in all self-consistent cases of Table \ref{tabas}.

The final result does not depend on $\gamma $, and on the self
-consistent
case studied, and has the form (valid for all positions ${\bf r}$)
\begin{eqnarray}
&&n\left( {\bf r},\infty \right) ={n\left( {\bf r},-\infty \right)
 \over 1+s n\left( {\bf r},-\infty \right) }, \nonumber
\\
&& \label{n}
\\
&&s={6\pi |g|{ } ^{2}\over \hbar \mu |\dot{B}|}={12\pi ^{2}\hbar
|a_{a}|\over m}{\Delta \over |\dot{B}|} . \nonumber
\end{eqnarray}
The product $s n$ in Eq.\ (\ref{n})  would be proportional to the
Landau-Zener exponent for the transition between the condensate and
the resonant molecular states whose energies cross due to the time
variation of the magnetic field, if one could keep the coupling
strength $g\varphi _{0}$  constant. However, for the non-linear curve
 crossing
problem represented by Eq.\ (\ref{GPE}), the Landau-Zener formula is
replaced by Eq.\ (\ref{n}), which predicts a lower crossing
probability, since the coupling strength $g\varphi _{0}$  decreases,
 along with
the decrease of the condensate density during the process.

The asymptotic result (\ref{n}) describes the decay of the
condensate {\it density.} Assuming a homogeneous initial density
 within the
trap, Eq.\ (\ref{n}) applies also to the loss of the total {\it
 population}
$N\left( t\right) =\int n\left( {\bf r},t\right) d^{3}r$.

An asymptotic expression for the total population can also
be found when the homogeneous distribution is replaced by the
Thomas-Fermi one (see Ref.\ \cite{MADKDK96}). In this case, given
$n_{0}$  is the maximum initial density in the center of the trap, one
obtains
\begin{eqnarray}
{N\left( \infty \right) \over N\left( -\infty \right) }={15\over 2sn{
 } _{0}}\biggl\{{1\over 3}+{1\over sn{ } _{0}}-{1\over 2sn{ }
 _{0}}\sqrt{1+{1\over sn{ } _{0}}} \nonumber
\\
\times \ln \biggl \lbrack \left( \sqrt{1+{1\over sn{ } _{0}}}+1\right
) /\left( \sqrt{1+{1\over sn{ } _{0}}}-1\right) \biggr \rbrack
 \biggr\} . \label{gd}
\end{eqnarray}
\section{Inclusion of loss by excitation}\label{SecExSt}

The other mechanism of molecular condensate loss, considered in
Ref.\ \cite{MJT99}, involves the decay of the resonant molecular
state by transferring atoms to excited (discrete) trap states or to
higher-lying non-trapped (continuum) states. This decay is possible
when the potential energy of the resonant state $V_{m}-\mu _{m}B\left
( t\right) $ exceeds the
energy of the atom pair formed. A simpler version of this mechanism
has been studied in Ref.\ \cite{AV99}.

Let us consider, for a moment, the decay of the molecule as a
``half collision'', ignoring the finite size of the trap. The excess
energy $E$ of the released pair, where $E=E\left( t\right) =\mu
 B\left( t\right) -V$ at the time of
release, determines a ``half-width'' of the Feshbach resonance.
According to Eq.\ (27) of Ref.\ \cite{MJT99} this energy-dependent
half-width is given by
\begin{equation}
\Gamma _{\text{cr}}\left( E\right) ={|a_{a}|\mu \Delta \over \hbar {
 } ^{2}}\sqrt{mE} . \label{gammacr}
\end{equation}
It should be recalled that here always $E>0$. The coupling of the
resonant molecular state with an excited state $v$ of an atom pair in
the trap can be described by a  coefficient $g_{v}$. This coefficient
 can
be related to $\Gamma _{\text{cr}}$  through (see Ref.\ \cite{MJT99})
\begin{equation}
g^{2}_{v}={\hbar \over \pi }\Gamma _{\text{cr}}\left( \epsilon
 _{v}\right)  {\partial \epsilon { } _{v}\over \partial v} ,
\end{equation}
where $\epsilon _{v}$  is the pair excited-state energy measured from
 the
ground trap state and $\partial \epsilon _{v}/\partial v$  measures
 the distance between the trap
states in the vicinity of $v$ (i. e., the inverse density of states).

As the magnetic field $B\left( t\right) =B_{0}+\dot{B}t$ rises above
 the resonant value
$B_{0}$, a crossing starts to occur between the resonant molecular
 state
and the excited trap levels. Neglecting molecular collisions and
motion, the resonant-state wavefunction $\varphi _{m}\left( t\right)
 $ is propagated from
$\varphi _{m}\left( t_{0}\right) $ according to the rule (see Ref.\
 \cite{YB98})
\begin{equation}
\varphi _{m}\left( t\right) =\varphi _{m}\left( t_{0}\right)
 \exp\left( -{\pi \over \hbar \mu \dot{B}}\sum\limits^{}_{v}|g_{v}
|^{2}\right)  . \label{LZsum}
\end{equation}
Although $g_{v}$, for a given $v$ state, is time-independent, the sum
taken over all $v$ states, bounded by the interval $\mu \dot{B}t_{0}
<\epsilon _{v}<\mu \dot{B}t$, is
time-dependent. Whenever the amplitude of each crossing is small,
i.e., when
\begin{equation}
|g_{v}|^{2}/\left( \hbar \mu \dot{B}\right) \ll 1 , \label{contcond}
\end{equation}
the sum in Eq.\ (\ref{LZsum}) can be replaced by an integral,
giving
\begin{equation}
\varphi _{m}\left( t\right) =\varphi _{m}\left( t_{0}\right)
 \exp\left( -{\pi \over \hbar \mu \dot{B}}\int\limits^{t}_{t{ } _{0}}
|g_{v}|^{2}{\mu \dot{B}\over \partial \epsilon _{v}/\partial
 v}dt\right)  , \label{LZint}
\end{equation}
where $\left( \partial \epsilon _{v}/\partial v\right) /\mu \dot{B}$
 measures the time interval between sequential
crossings. Differentiation of Eq.\ (\ref{LZint}) with respect to $t$
gives the following expression
\begin{equation}
{\partial \varphi { } _{m}\over \partial t}=-{\pi \over \hbar }|g_{v}
|^{2}{\partial v\over \partial \epsilon { } _{v}}\varphi _{m}\equiv
-\Gamma _{\text{cr}}\left( \epsilon _{v}\right) \varphi _{m} .
\end{equation}
It should be kept in mind that this result is valid only in cases
in which the energy gap between the atomic and molecular states
increases rather fast; i.e., when $\mu \dot{B}>0$ and the condition
(\ref{contcond}) is obeyed. In the case of $\mu \dot{B}<0$,
 transitions from the
ground trap state to excited ones are counterintuitive (see Ref.\
\cite{YBJB99}), and become negligible when
\begin{equation}
\delta B\gg {g\sqrt{n}g{ } _{v}\over \hbar \mu \omega { }
 _{\text{trap}}} , \label{NDcond}
\end{equation}
where $\delta B$ is the range of variation of $B$ extended over both
 sides
of the resonance.

Thus, whenever the energy gap between the resonant molecular and
condensate states is positive, it increases rather fast [following
Eq.\ (\ref{contcond})], and Eq.\ (\ref{NDcond}) is obeyed, one can
account for the decay of the resonant molecular state into excited
trap states by adding a term $-i\hbar \Gamma _{\text{cr}}\left( \mu
 \dot{B}t\right) \varphi _{m}$  to the right hand side of
Eq.\ (\ref{GPEm}) or by substituting $\Gamma _{m}=\gamma n+\gamma
 _{m}n_{m}+\Gamma _{\text{cr}}\left( \mu \dot{B}t\right) $ in Eqs.\
(\ref{deneq}). In this way, the two mechanisms can be combined in a
single formalism. Calculations made using this formalism are discussed
in the following section.

\section{Results and Discussion}\label{SecDisc}

Calculations have been carried out on the loss of atoms for the
case of the Na BEC experiments \cite{IASMSK98,SIAMSK99} using both
analytical results [Eqs.\ (\ref{K3}), (\ref{nt}), (\ref{n}), and
(\ref{gd})] and numerical solutions of Eq.\ (\ref{deneq}). The
parameters used to describe the system, reported previously
\cite{MJT99,AV99}, are $a_a$ = 3.4nm, $\mu = 2\mu_a-\mu_m =
3.65\mu_B$ (where $\mu_B = 9.27\times 10^{-24}$J/T is the Bohr
magneton), and $\Delta =$ 0.95$\mu$T and 98$\mu$T, respectively, for
the two resonances observed at 85.3mT (853G) and 90.7mT
(907G)~\cite{IASMSK98,SIAMSK99}.  These values of $\Delta$ agree
with the measured value for the 90.7mT resonance \cite{SIAMSK99},
and with the indirectly inferred order of magnitude for the 85.3mT
resonance \cite{SIAMSK99}. The scattering lengths $a_{m}$  and
 $a_{am}$  for
molecule-molecule and atom-molecule collisions, respectively, are
not known, but calculations show that the results of our analysis
are practically insensitive to the variation of their values as long
as they stay within the order of magnitude of $a_{a}$.

In the calculations one should make a clear distinction between
the two types of experiments conducted at MIT --- the slow-sweep and
the fast-sweep experiments. In the first case, the values of the
magnetic field at which the ramp stops short of resonance are such
that the conditions needed for the excitation mechanism to occur do
not exist, and only the collisional deactivation applies.

\begin{figure}
\newpage
\psfig{clip=,figure=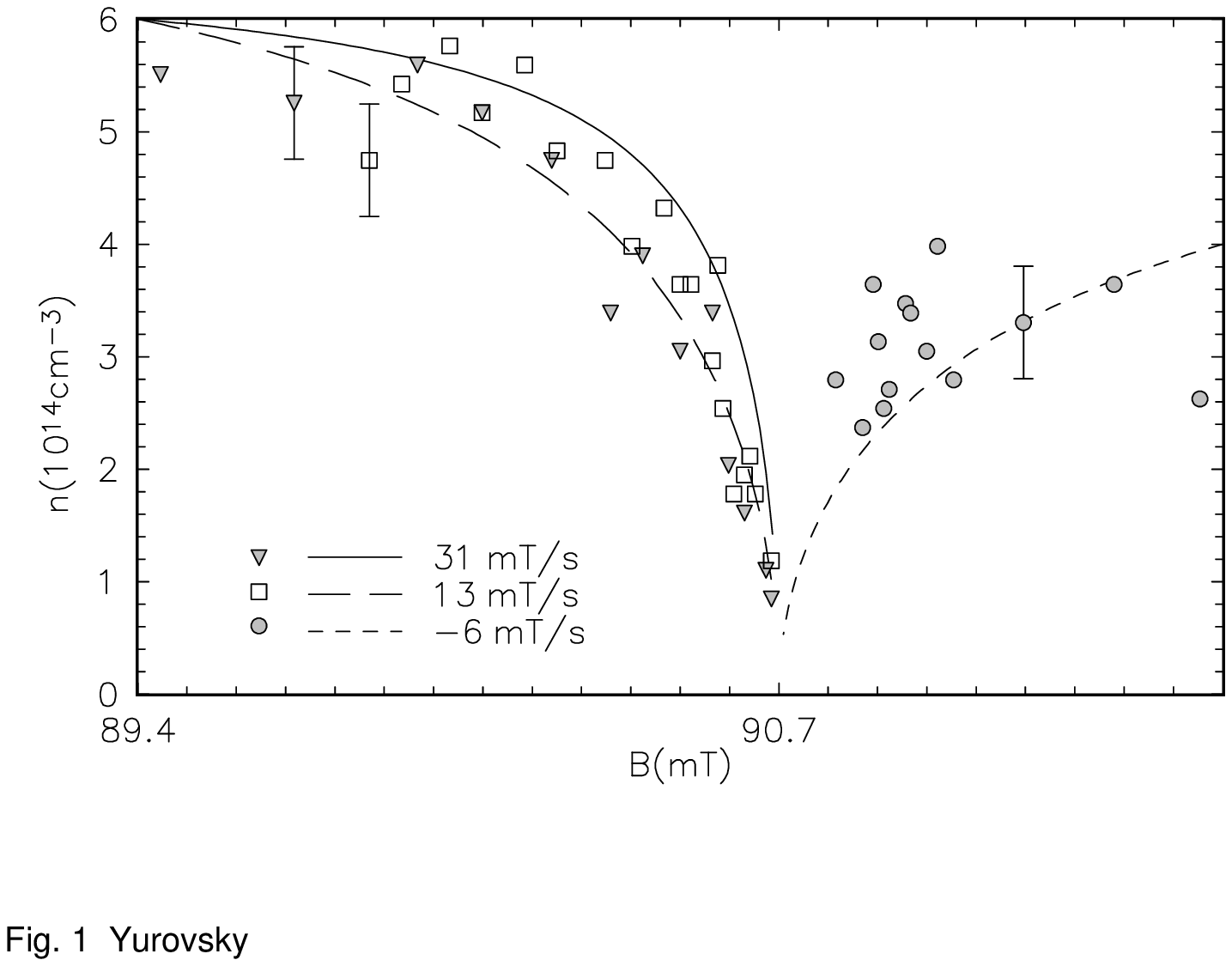,width=3.375in}

\caption{The surviving mean density vs. the stopping value of the
magnetic field, calculated with the optimal value of the deactivation
parameter $\gamma =0.8\times 10^{-11}$  cm$^{3}/$s. The resonance was
 approached from below with
two ramp speeds, 13 mT/s and 31 mT/s, or from above, with the ramp
 speed -6
mT/s. These are compared with the experimental results
\protect\cite{SIAMSK99} (squares, triangles and circles) for several
 values
of the ramp speed $dB/dt$ (in mT/s). \label{fig_n}}

\psfig{clip=,figure=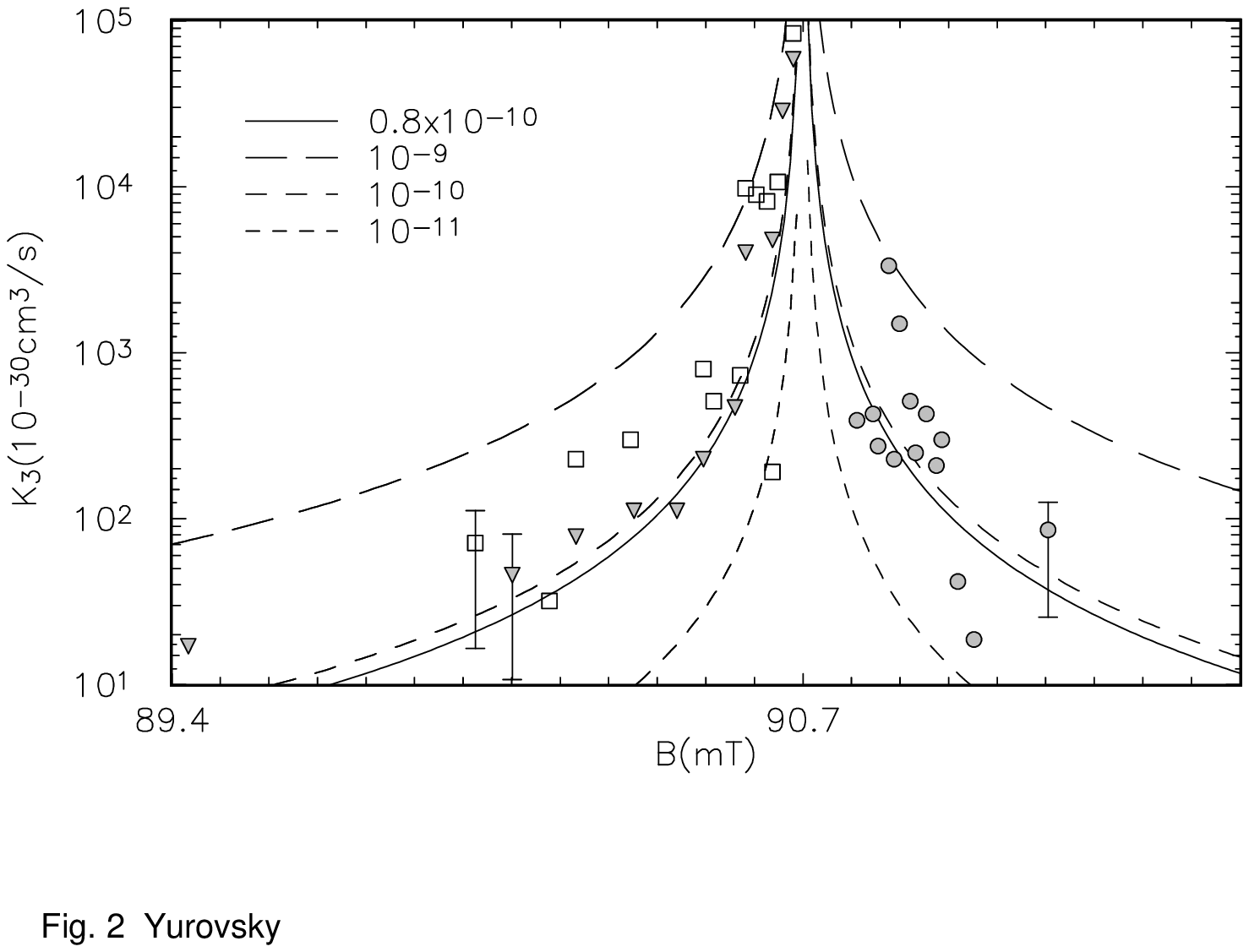,width=3.375in}

\caption{The 3-body rate coefficient ($K_{3}$) vs. the stopping value
of the magnetic field, calculated  with the optimal and other values
 of the
deactivation parameter $\gamma $ (in units of cm$^{3}$/s), on
 approaching
the resonance from below or above. The calculated value of $K_{3}$  is
independent of the ramp speed value. Other notations as in Fig.\ 1.
\label{fig_K3}}

\newpage
\psfig{clip=,figure=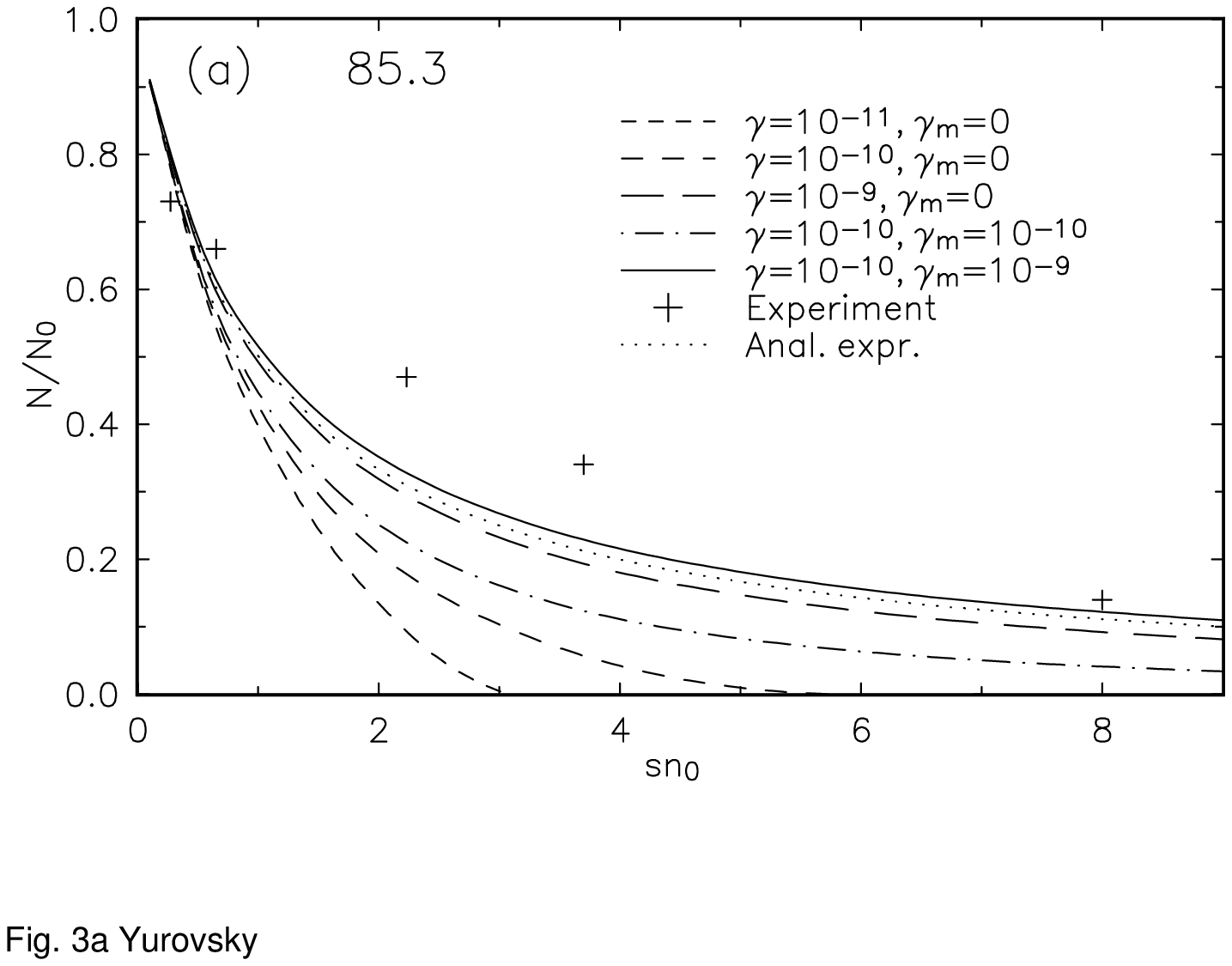,width=3.375in}
\psfig{clip=,figure=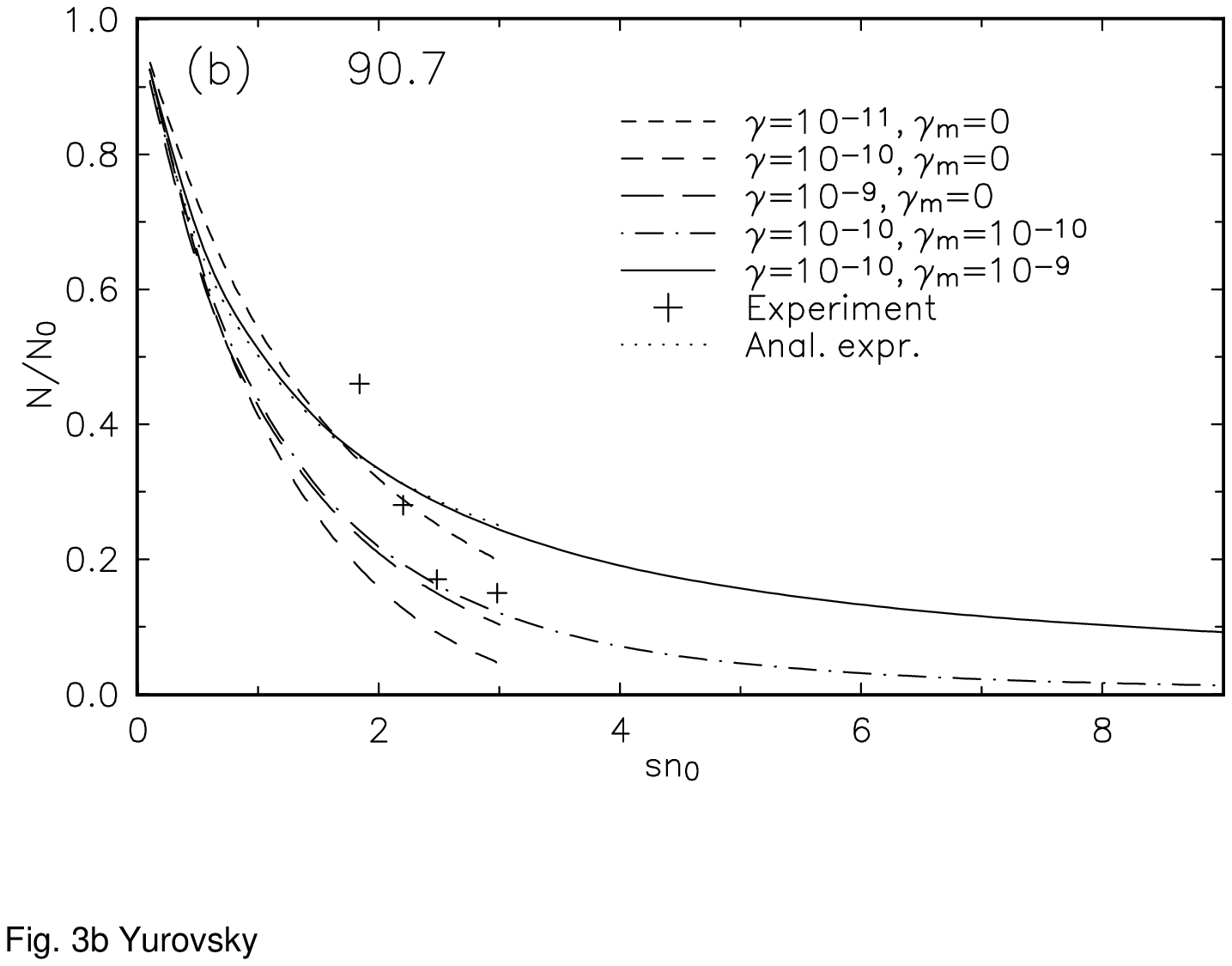,width=3.375in}

\caption{Ratio of surviving trap population $N$ to the initial one
$N_{0}$ for the 85.3mT (853G) and 90.7mT (907G) resonances [parts (a)
 and
(b), respectively] in the  homogeneous-density approximation vs. $s
 n_{0}$
(where the parameter $s$ is defined by Eq.\ (\protect\ref{n}) and
 $n_{0}$
is the initial density). The curves show results of calculations
 carried
out for different magnitudes of the parameters $\gamma $ and $\gamma
 _{m}$  (in units
of cm$^{3}$/s), without accounting for the excitation mechanism. The
asymptotic analytical result Eq.\ (\protect\ref{n}) is given by the
 dotted
line. The results of the MIT fast-sweep experiment (Ref.\
\protect\cite{SIAMSK99}) are shown for comparison.
 \label{fig_loss_da}}

\newpage
\psfig{clip=,figure=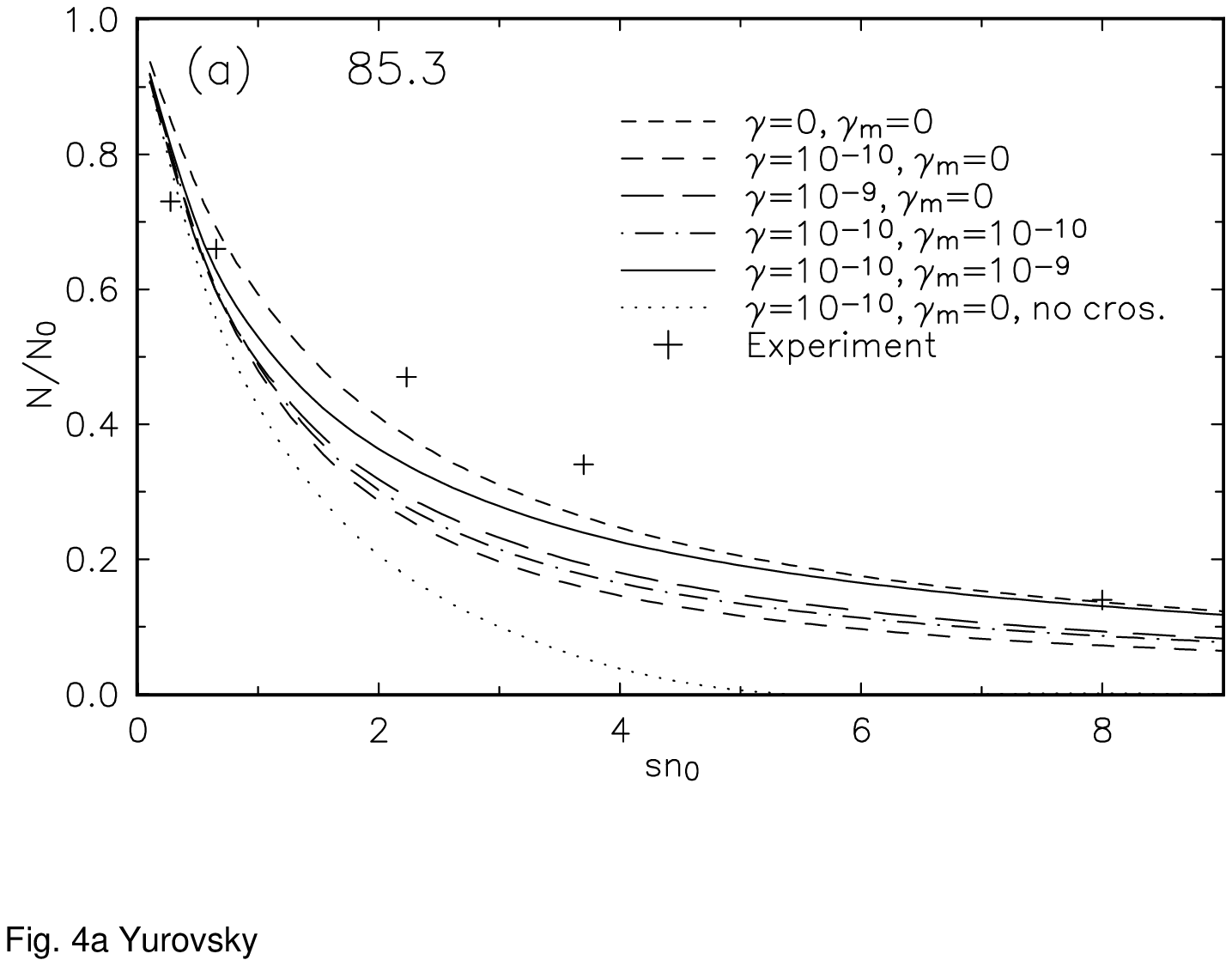,width=3.375in}
\psfig{clip=,figure=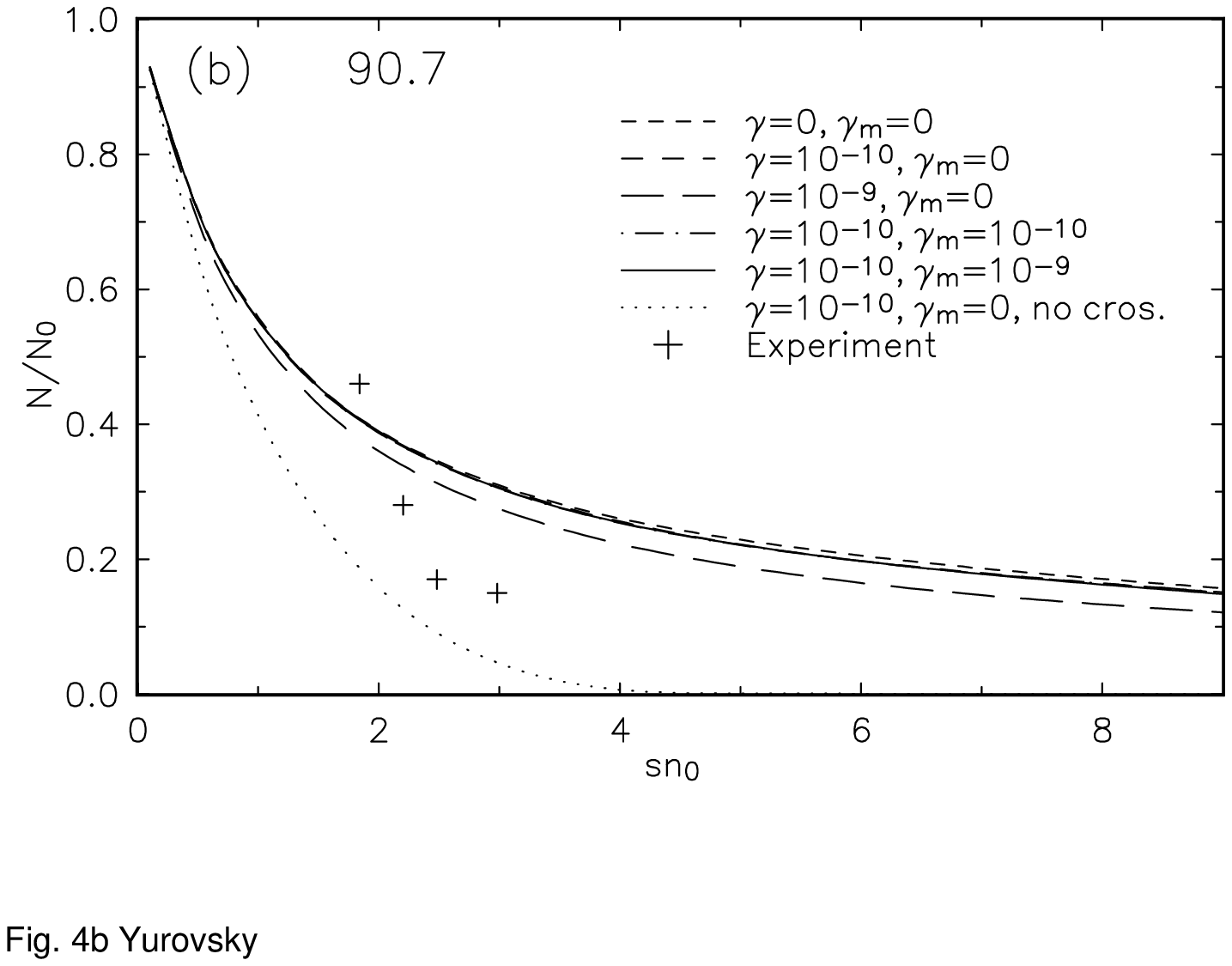,width=3.375in}

\caption{Same as Fig.\ \protect\ref{fig_loss_da} but accounting for
the excitation mechanism. The results of calculations without
 accounting
for the excitation mechanism are given for comparison by the dotted
 line.
In the part (b) the plot calculated for $\gamma =10^{-10}$cm$^{3}/$s
 and different
magnitudes of $\gamma _{m}$  are practically
 indistinguishable.\label{fig_loss_ex}}

\newpage
\psfig{clip=,figure=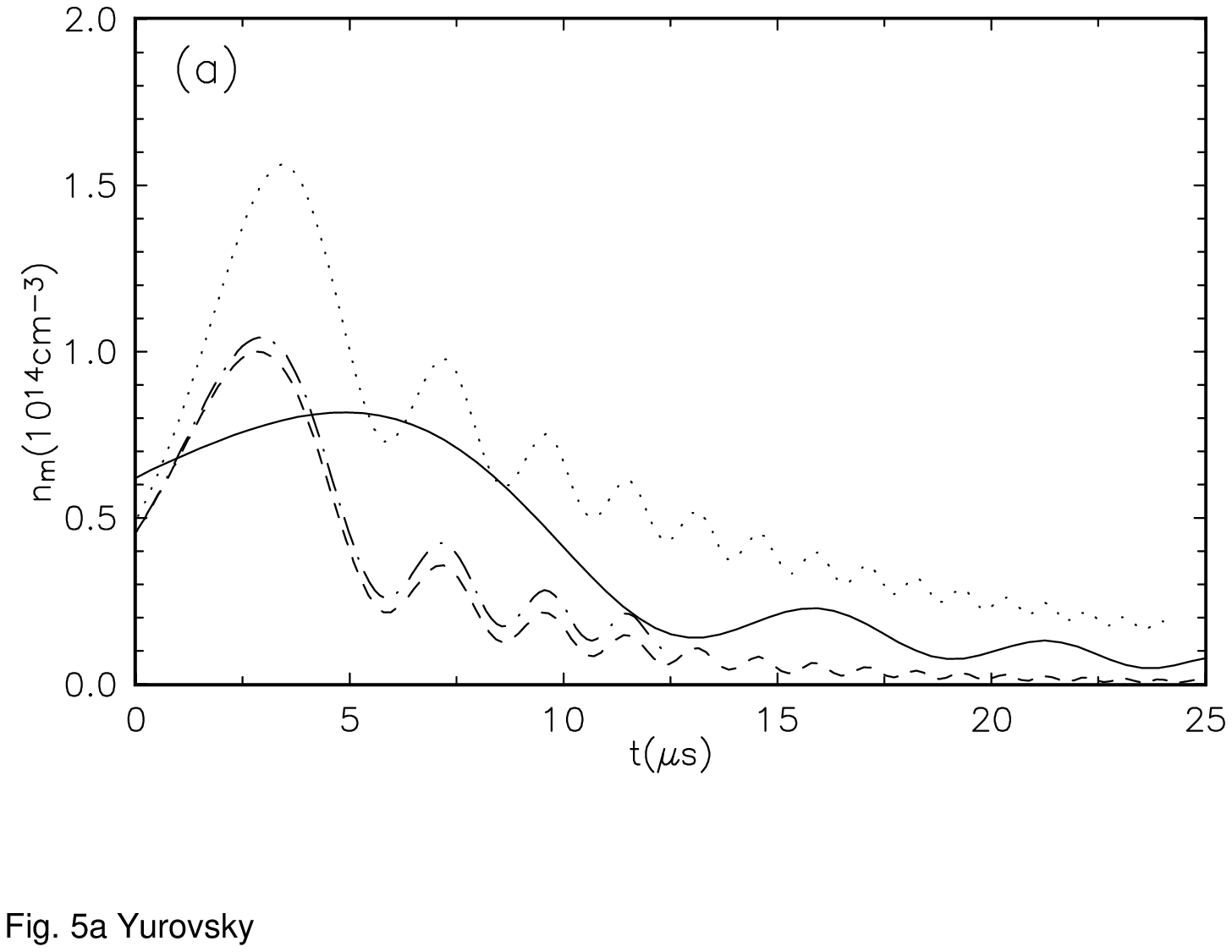,width=3.375in}
\psfig{clip=,figure=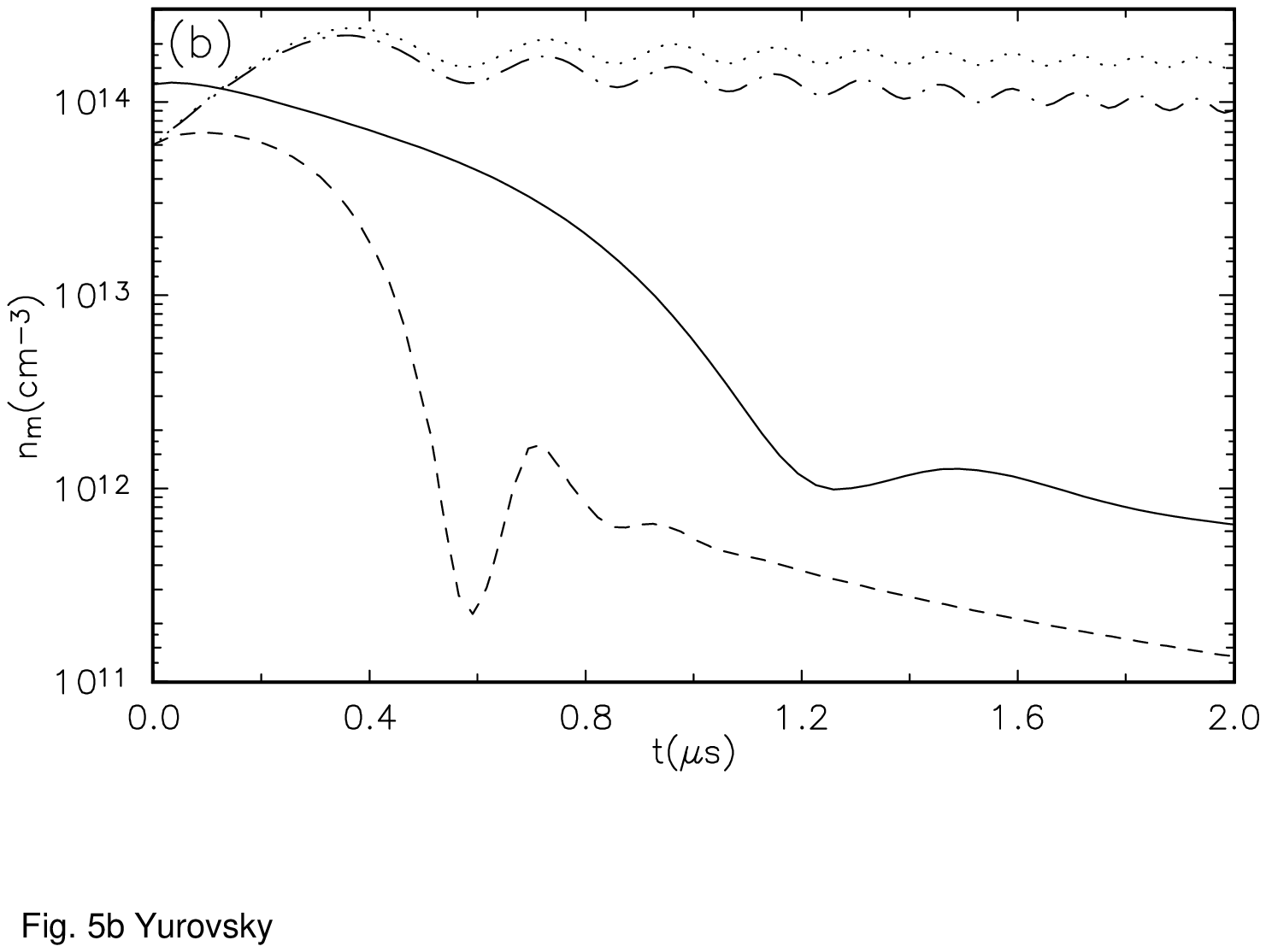,width=3.375in}

\caption{Time-dependence of the molecular condensate density in
cm$^{-3}$ during fast passage through the 85.3mT (853G) and 90.7mT
(907G)
resonances [parts (a) and (b), respectively], calculated using the
 value of
$\gamma =10^{-10}$cm$^{3}/$s. The dashed and solid lines are
 calculated accounting for the
combined effect of the two mechanisms (deactivation and excitation)
 for
$sn_{0}=1$ and 5, respectively, and $\gamma _{m}=10^{-9}$cm$^{3}/$s.
 The dotted and dash-dotted
lines refer only to the deactivation mechanisms for $\gamma _{m}=0$
 and $10^{-9}$cm$^{3}/$s,
respectively, and $sn_{0}=1$. The oscillations of the molecular
 density are
faster for the lower value of $sn_{0}$. \label{fig_td}}

\newpage
\psfig{clip=,figure=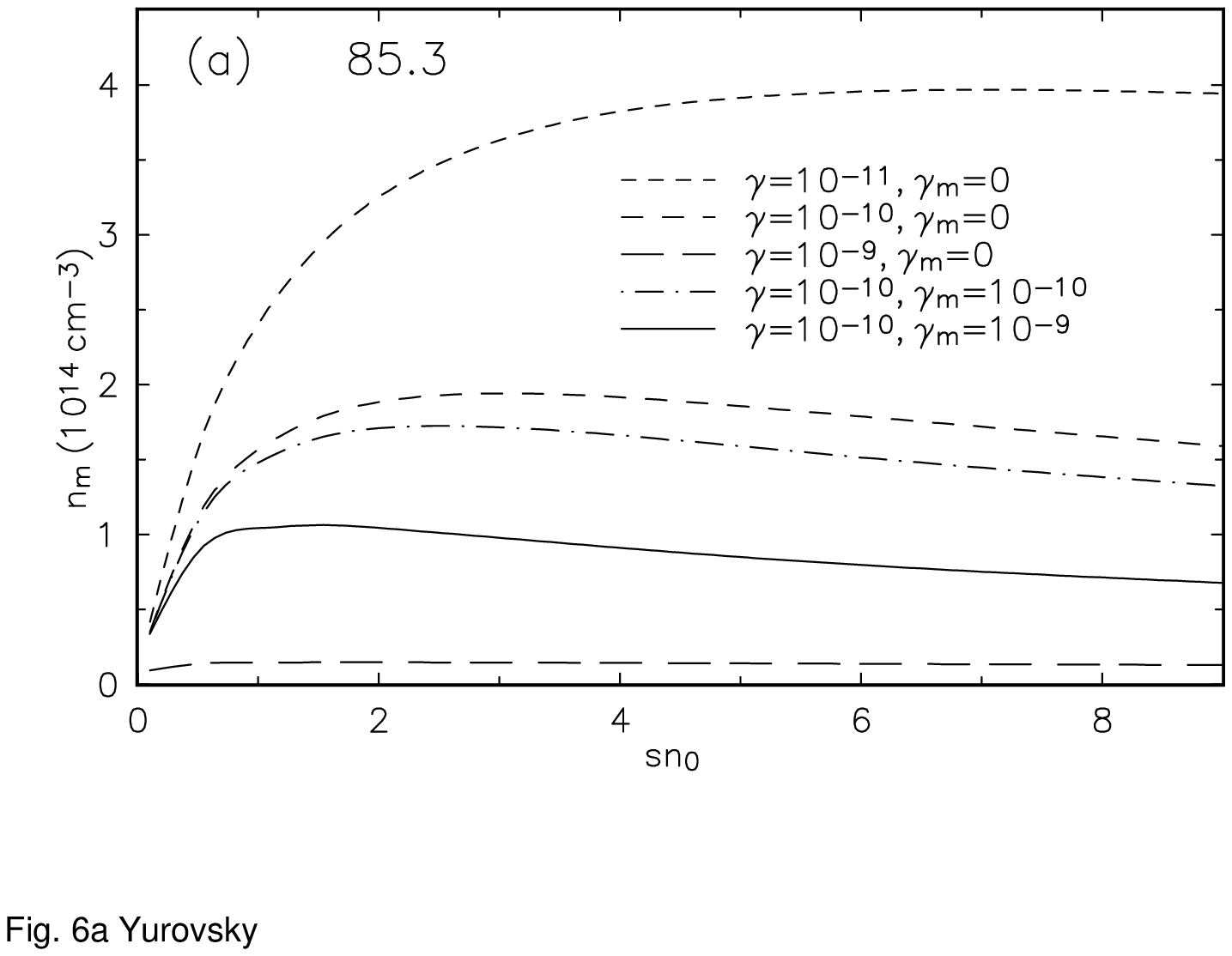,width=3.375in}
\psfig{clip=,figure=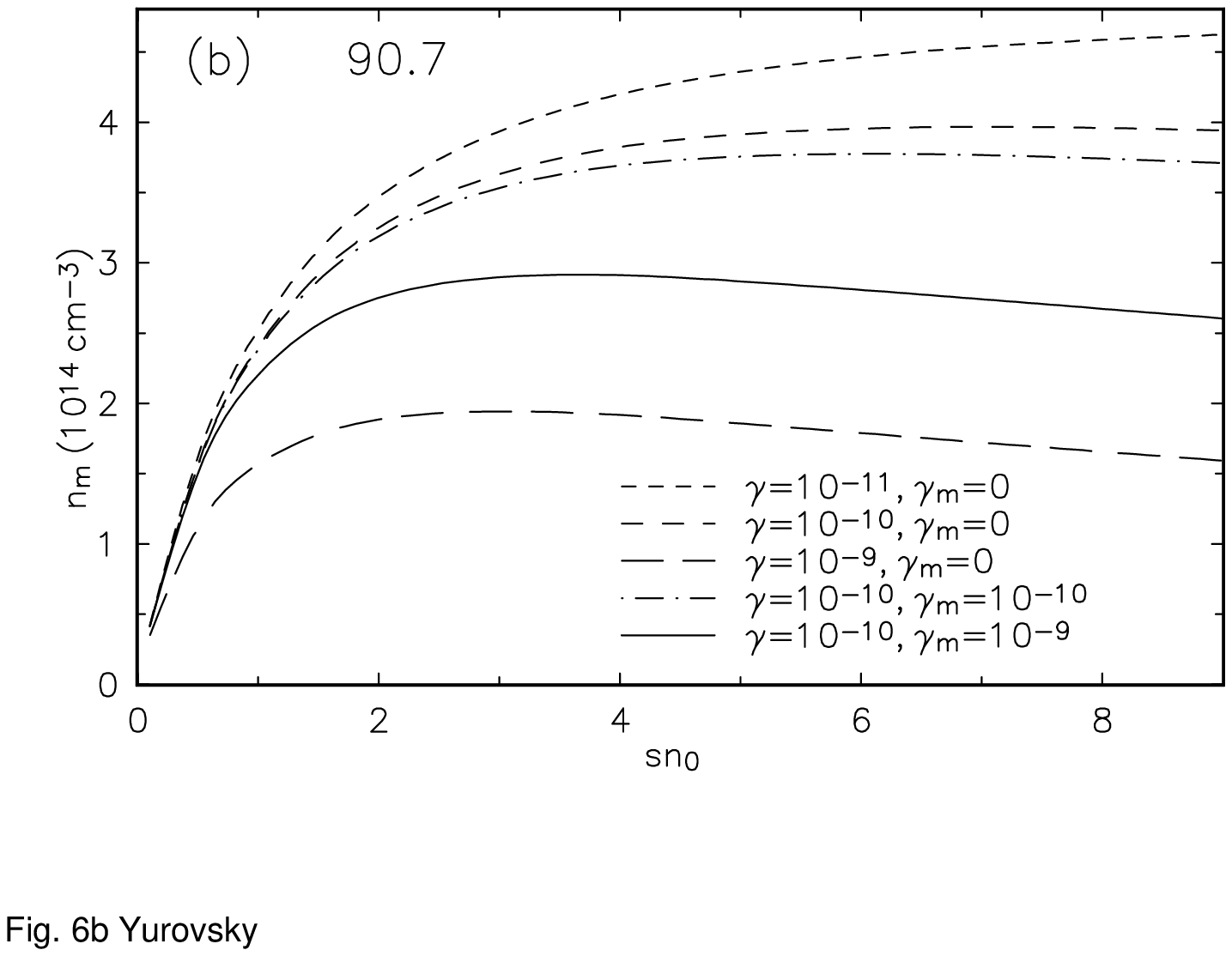,width=3.375in}

\caption{Peak values of the molecular condensate density in cm$^{-3}$
attained during fast passage through the 85.3mT (853G) and 90.7mT
(907G)
resonances [parts (a) and (b), respectively] vs. $sn_0$ for various
magnitudes of the  parameter$s \gamma$ and $\gamma_{m}$ (see Caption
 of
Fig.\ \protect\ref{fig_loss_da}), calculated without accounting for
 the
excitation mechanism.\label{fig_den_da}}

\newpage
\psfig{clip=,figure=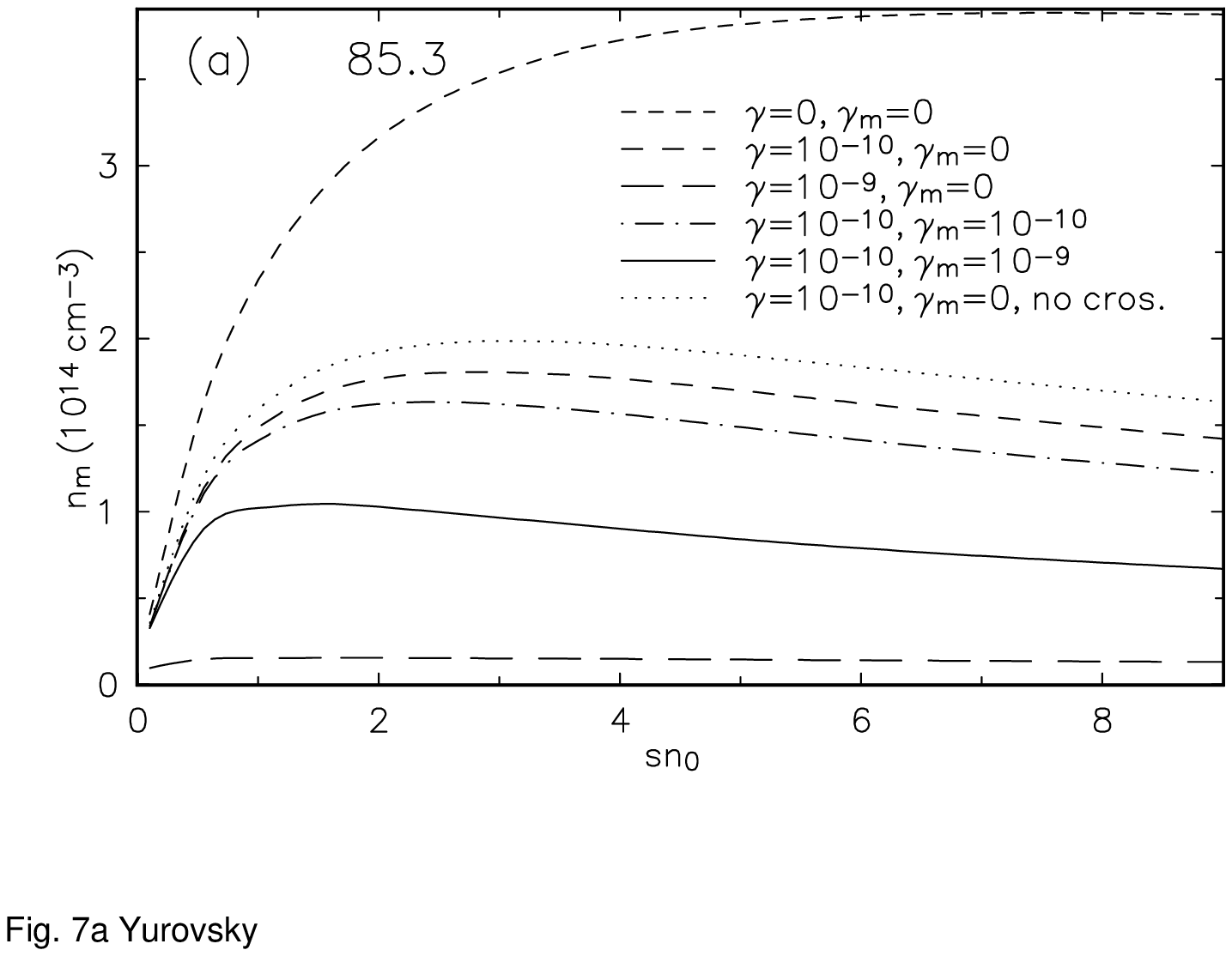,width=3.375in}
\psfig{clip=,figure=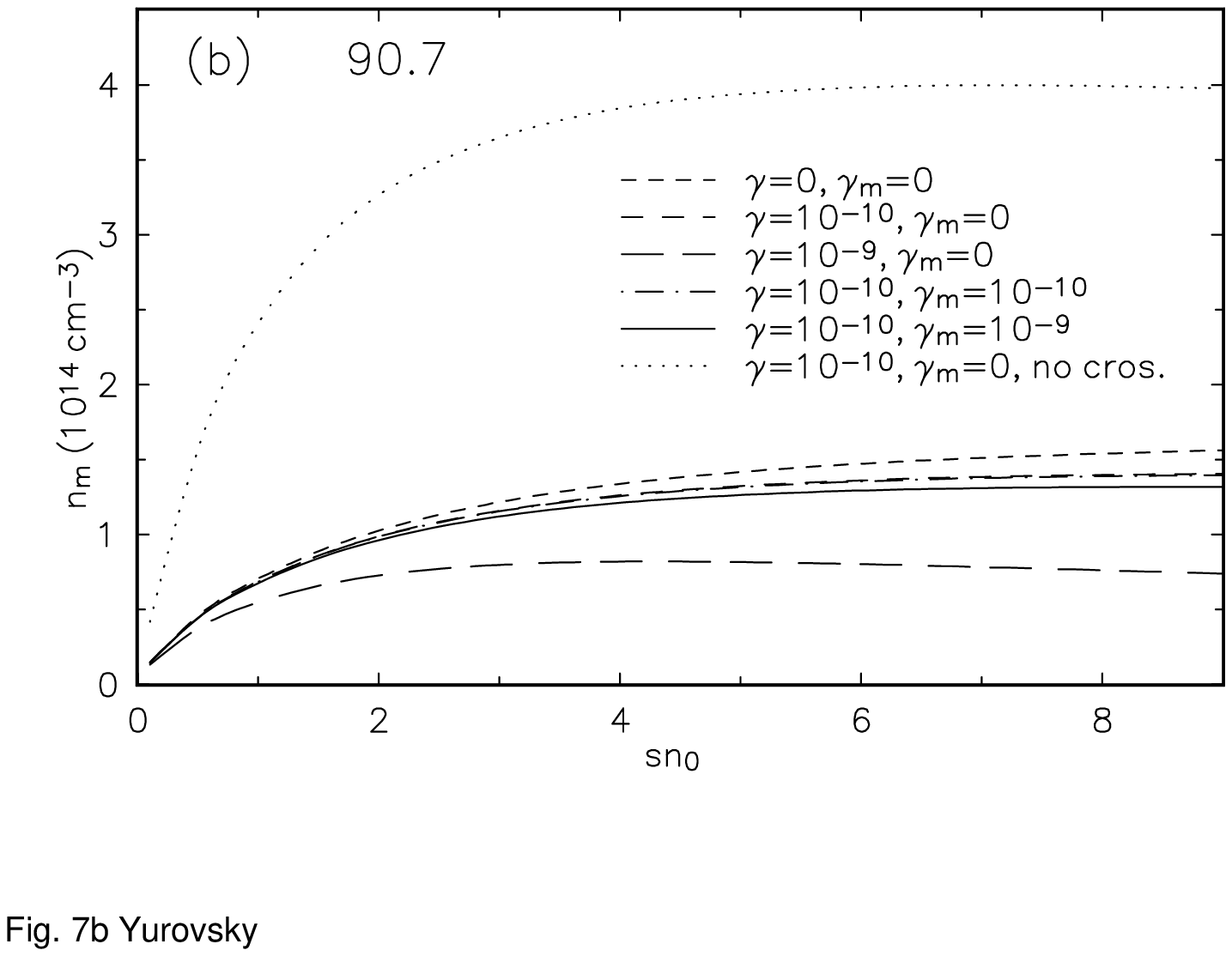,width=3.375in}

\caption{Same as Fig.\ \protect\ref{fig_den_da} but accounting for the
excitation mechanism. The results of calculations without accounting
 for
the excitation mechanism are given for comparison by the dotted line.
 In
the part (b) the plot calculated for $\gamma =10^{-10}$cm$^{3}/$s
 with $\gamma _{m}=0$ and $10^{-10}$
cm$^{3}/$s  are practically indistinguishable.\label{fig_den_ex}}

\end{figure}

The graphs shown in Figs.\ \ref{fig_n} and \ref{fig_K3} pertain
to the slow-sweep MIT experiment for the strong 90.7 mT resonance.
This resonance has been approached from below with two ramp speeds,
and from above with one. Figure \ref{fig_n} shows the surviving atomic
density $n$, and Fig.\ \ref{fig_K3} shows $K_3$, both vs. the stopping
value of the magnetic field $B$. The difference between Eqs.\
(\ref{K3}), (\ref{nt}) and the results of a direct numerical solution
of Eqs.\ (\ref{deneq}) for all ramp speeds is so small that the
corresponding plots are indistinguishable. The calculated plots were
obtained using homogeneous-density initial conditions, starting from a
$B$ value of 89.4mT on approach from below and 91.6mT from above. The
corresponding initial mean densities were extracted from the
experimental data \cite{SIAMSK99}. A best fit of the parameter
 $\gamma $,
using Eq.\ (\ref{nt}), to the MIT data gives $\gamma =0.8\times 10^{
-10}$  cm$^{3}/$s (see
Fig.\ \ref{fig_n}). But owing to the large scattering of the
experimental data, and the associated uncertainty in the value of
 $\gamma $,
we proceeded using the value of $10^{-10}$cm$^{3}/$s in our
calculations, following Ref.\ \cite{YBJW99}. Given a density of about
$10^{15}$cm$^{-3}$ this value of $\gamma $ implies a deactivation
 time of
$\sim 10^{-5}$s. The molecule-molecule deactivating collisions are
negligible compared to the atom-molecule collisions due to the small
molecular density [see Eq.\ (\ref{quasist})] whenever the fast decay
condition (\ref{FDA}) holds.

The rate of condensate loss due to atom-molecule deactivating
collisions was also studied in Ref.\ \cite{TTHK99}. As noted in the
previous section, the expression for $K_{3}$  obtained there is
 smaller by
a factor 1.5 from that of Eq.\ (\ref{K3}) [see discussion after this
equation]. Without this factor one would not obtain the almost perfect
match between the analytical and the numerical results mentioned in
the discussion of Figs.\ \ref{fig_n} and \ref{fig_K3} above. This
omission has been corrected in Ref.\ \cite{AV99} and the value they
obtain for their $G_{\text{stab}}$(corresponding to our $2\gamma $),
 of $4\times 10^{-10}$  cm$^{3}/$s,
was estimated by best agreement with the experimental data
\cite{SIAMSK99}. This value is 2.5 times bigger than our estimate.
This discrepancy may be due to the large scatter in the experimental
data. The estimate of Ref.\ \cite{AV99} is based only on the
experimental $K_{3}$  plot (see Fig.\ \ref{fig_K3}) which is obtained
 by a
differentiation of the experimental density data, and therefore shows
a scatter of the data of up to an order of magnitude (see Fig.\
\ref{fig_K3}, as well as the corresponding figure in Ref.\
\cite{SIAMSK99}, that use a logarithmic scale). Equation (\ref{nt})
(presented in Ref.\ \cite{YBJW99}) allows us to estimate the value of
$\gamma $ directly from the experimental plot for the atomic density
(see
Fig.\ \ref{fig_n}), that shows a much smaller ($\approx 20\%$)
 scatter of the
data points. Anyhow, in both cases, the error bar should be comparable
to the value of $\gamma $ itself.

An inelastic rate coefficient $2\gamma$ with a magnitude of the
order of $10^{-10}$cm$^{3}/$s appears to be reasonable.  First, this
value is two orders of magnitude smaller than the upper bound set by
the unitarity constraint on the $S$-matrix~\cite{MJ89}.  In the limit
of small momentum, unitarity provides $2\gamma \le \hbar \lambda /m$,
where $\lambda $ (the de Broglie wavelength) in the current situation
is limited by the experimental trap dimensions.  This constraint sets
an upper bound of $2.5\times 10^{-8}$cm$^{3}/$s to $2\gamma$. Second,
our estimate of $10^{-10}$cm$^{3}/$s for $\gamma$ is consistent with
the order of magnitude of recently calculated~\cite{Forrey99}
vibrational deactivation rate coefficients due to ultracold collisions
of He with H$_{2}$ in highly excited vibrational levels.

The remaining figures (\ref{fig_loss_da} to \ref{fig_den_ex})
pertain to the fast-sweep MIT experiment. Figures \ref{fig_loss_da}
and \ref{fig_loss_ex} present the surviving part of the trap
population after passing through each of the two resonances. Following
the experimental conditions, the value of $10^{15}$  cm$^{-3}$  is
 used for the
initial density, and the magnetic field starting and stopping values
are shifted from resonance by 3 mT for the strong (90.7 mT) resonance
and by 2 mT for the weak (85.3 mT) one.

The analytical results of Eq.\ (\ref{n}), together with the
direct numerical solutions of Eqs.\ (\ref{deneq}) for the homogeneous
initial distribution, considering only the deactivating mechanism, are
compared in Fig.\ \ref{fig_loss_da} with the results of the fast-sweep
experiment \cite{IASMSK98,SIAMSK99}. Although Fig.\ \ref{fig_loss_da}
does not specify the direction of the Zeeman shift $\mu \dot{B}$, one
 should
recall (see Secs.\ \ref{Introd} and \ref{PasRes} above) that the
excitation mechanism can be ignored when $\mu \dot{B}<0$. The
 asymptotic result
(\ref{n}) reproduces a characteristic dependence on the ramp speed,
although it is independent of $\gamma $. The numerical solutions
 clearly show
a dependence on $\gamma $, as assumptions (\ref{FDA})  and
(\ref{ascond})
underlying the asymptotic result (\ref{n}) do not hold in this case.
The loss reaches a maximum at a value of $\gamma $ dependent on the
 various
parameters (e.g., $\gamma \approx 10^{-10}\text{cm}^3/\text{s}$ for
the 90.7 mT resonance when $s n_0 \approx 2$), as a result of the
conflicting asymptotic and fast-decay conditions (\ref{FDA})  and
(\ref{ascond}). The calculated drop in the condensate loss on increase
of the molecular dumping rate $\gamma _{m}$, which may seem
 paradoxical, can have
the following explanation. Reaction (\ref{SCol}) leads to the loss of
three condensate atoms per each resonant molecule formed, while
reaction (\ref{MCol}) leads to the loss of only two condensate atoms.
In the present case, the loss rate is limited by resonant molecule
formation [reaction (\ref{RCol})]. Therefore, the increase of $\gamma
 _{m}$
transfers flow from channel (\ref{SCol}) to channel (\ref{MCol}), thus
reducing the number of condensate atoms lost per each resonant
molecule formed.

The results of the calculations, in which the effect of crossing
to  excited trap states is incorporated, are plotted in Fig.\
\ref{fig_loss_ex}. The crossing rate was calculated with Eq.\
(\ref{gammacr}), using the values of $\Delta $, $a_{a}$, and $\mu $
 given at the
beginning of this Section. The results clearly show that the two
mechanisms do not augment each other. Adding the two-body excitation
mechanism to the more efficient three-body deactivation mechanism
actually reduces the loss. A possible explanation of this paradoxical
result is similar to the one used in explaining Fig.\
\ref{fig_loss_da}.

Our predicted losses are somewhat higher than the ones obtained
in the experiments for the 85.3 mT resonance, but significantly lower
for the 90.7 mT resonance. Actually, the closest we can get to the
experimental results is by considering only the two-body excitation
mechanism for the 85.3 mT resonance, and the three-body deactivation
mechanism for the 90.7 mT resonance.

The results of Ref.\ \cite{AV99}, considering only the two-body
mechanism, also show a better agreement for the 85.3 mT resonance.
There are, however, significant differences between the theory used
there and the one presented here and in Ref.\ \cite{MJT99}. In Ref.\
\cite{AV99} the parameter $\gamma _{0}$  (analogous to our $\Gamma
 _{\text{cr}}$) is considered as
a constant, independent of the released energy $E$ or time $t$, and
prevailing all along the sweep, below and above the resonance. The
time integral in their Eq.\ (4) should be smaller by a factor of 2 if
the correct energy dependence of $\gamma _{0}$  is used. In our
 version,
following Ref.\ \cite{MJT99}, $\Gamma _{\text{cr}}$  is $t$
-dependent, exists only for
$E>0$, and attains the correct Wigner limit when $E\rightarrow 0$.
 The final result
of Ref.\ \cite{AV99} is an expression similar to our Eq.\ (\ref{n}),
with the exception that our parameter $s$ is larger by a factor $3/2$
 than
the corresponding parameter in Ref.\ \cite{AV99}. This difference
reflects the fact that Eq.\ (\ref{n}) was derived for the three-body
deactivation process, while Ref.\ \cite{AV99} deals exclusively with a
two-body excitation process. As a matter of fact, their coefficient
should be further reduced by a factor of 2 associated with the energy
dependence discussed above. Therefore the mechanism of Ref.\
\cite{AV99} requires a factor $s$ 3 times smaller than the one used in
Eq.\ (\ref{n}) for the deactivation mechanism.

Our results also differ from those of Ref.\ \cite{MJT99}. The
present calculations take into account the decrease of the condensate
density during the crossing [see discussion after Eq.\ (\ref{n})] and
therefore produce a lower condensate loss than the one obtained in
Ref.\ \cite{MJT99}, in which the loss is described by a Landau-Zener
formula. In the limit of a small loss, Eq.\ (\ref{n}) may resemble a
Landau-Zener expression in which $s n$ is substituted for the
 exponent.
However, a study of the associated free-atom problem shows that, even
when we retain the initial density $n\left( {\bf r},-\infty \right) $
 in the exponent, the latter
would still be smaller than $s n$ by a factor of 3. This factor is
directly related to the many-body character of the Gross-Pitaevskii
equations.

At the first stage of the atomic condensate loss (\ref{RCol}) a
condensate of molecules in the resonant state is formed. This
molecular condensate is unstable, due to the deactivation of the
molecules by reactions (\ref{SCol}) and (\ref{MCol}), as well as their
decay through the excitation mechanism. Figure \ref{fig_td} presents
the calculated time dependence of the molecular condensate density,
for various magnetic field ramp speeds, taking into account the
various loss mechanisms. The oscillations of the molecular condensate
density are connected to the intercondensate tunneling considered in
Ref.\ \cite{TTHK99}. Figure \ref{fig_td} shows that the excitation
loss enhances the damping of the oscillations and the decay of the
molecular density. Nevertheless, the molecular condensate persists at
least a few tenths of a microsecond before decaying. This time is long
enough to allow converting the population of the vibrationally-excited
state $m$ to the ground molecular state by methods of coherent control
\cite{SB97,GR97}. Various techniques exist today for transfering
populations coherently to a preselected state \cite{BTS98}. The choice
of technique would be dictated by properties of the process and of the
target state (such as selection rules and stability).

The calculated peak values of the resonant molecular state
density are presented in Figs.\ \ref{fig_den_da} and \ref{fig_den_ex}
for various rates of the deactivating collisions. Figure
\ref{fig_den_da} refers only to the deactivation mechanism involving
atom-molecule and molecule-molecule collisions, which take place for
$\mu \dot{B}<0$, whereas Fig.\ \ref{fig_den_ex} takes into
 consideration the
combined effect of the two mechanisms (deactivation and excitation)
which may take place for $\mu \dot{B}<0$. These figures show that
 from about 10\%
to 90\% of the atomic density can be converted temporarily to a
molecular condensate in the resonant state, in spite of losses due to
the excitation mechanism and the molecule-molecule deactivation
collisions. The calculated molecular condensate density is higher for
the 90.7 mT resonance, or when the magnetic field ramp speed is not
too large.

\section{Conclusions}

This paper discusses the two types of loss experiments
\cite{IASMSK98,SIAMSK99} conducted at MIT on sodium BEC, using a
time-varying magnetic field in the proximity of a Feshbach resonance.
The various processes discussed here involve a temporary formation of
a molecular condensate. In the slow-sweep experiment the dominant
loss mechanism is three-body deactivation by atom-molecule inelastic
collisions. A best fit of an analytical expression for the atomic
density obtained here [Eq.\ (\ref{nt})] to the MIT data yields a rate
coefficient for deactivating atom-molecule collisions of
$2\gamma=1.6\times 10^{-10}$  cm$^{3}/$s. For the fast-sweep
experiment an analytical expression, generalizing the Landau-Zener
formula to a case of coupled nonlinear equations, is obtained. A
different two-body mechanism, involving an excitation of the
condensate by curve crossing, has been proposed previously
\cite{MJT99}. The combined effect of the two mechanisms is studied
here, including also molecule-molecule deactivating collisions. The
analysis shows that both processes should be taken into account, that
they do not contribute additively to the loss, and that the outcome
of the competition between them varies from one Feshbach resonance to
another. Our numerical results show that, under favorable conditions,
a substantial fraction of the trap population is converted to an
unstable molecular condensate. This condensate persists long enough
to allow its coherent transfer to a more stable state.

%\newpage
\mediumtext
\begin{table}

\caption{Asymptotic conditions as consistency tests on
solutions of Eq.\ (\protect\ref{RE}) \label{tabas}}
\begin{tabular}{llll} &case&$\dot{n}$&$n$\\ \tableline
a&$n\rightarrow $const&$\dot{n}\rightarrow 0$&$n\rightarrow $const\\
b&$n/t\rightarrow 0$&$\dot{n}\sim -6|g|^{2}\gamma n^{3}/\left( \mu
 \dot{B}t\right) ^{2}$&$n\rightarrow \text{const}$\\
c&$n/t\rightarrow \infty $&$\dot{n}\sim -6|g|^{2}n/\left( \hbar
 ^{2}\gamma \right) $&$n\sim \exp\left\lbrack -6|g|^{2}t/\left( \hbar
 ^{2}\gamma \right) \right\rbrack $\\
d&$n/t\rightarrow $const&$\dot{n}\sim t$&$n\sim t^{2}$\\

\end{tabular}
\end{table}
\end{document}